\documentclass[letterpaper, 12pt]{article} 
\usepackage{geometry}
\usepackage{amsmath}
\usepackage{amsfonts}
\usepackage{amssymb}
\usepackage{graphicx}
\usepackage{hyperref}
\usepackage{setspace} 
\usepackage{fancyhdr} 
\usepackage{times}
\usepackage{booktabs}
\usepackage{float}

\usepackage{adjustbox}
\usepackage{siunitx}
\usepackage{subcaption}
\usepackage{longtable}
\usepackage{fancyhdr}
\usepackage{lscape}
\usepackage[sectionbib]{chapterbib}
\usepackage[super,comma]{natbib}
\makeatletter
  \renewcommand\l@section{\@dottedtocline{2}{1.5em}{3em}}
  \renewcommand\l@subsection{\@dottedtocline{2}{3em}{3em}}
\makeatother
\title{\centering {\fontsize{24pt}{24pt}\selectfont Human Trust in AI Search: \\A Large-Scale Experiment}}
\author{Haiwen Li $^{1}$ \and
Sinan Aral$^{1,2}$\thanks{Corresponding Author (Sinan Aral; \href{mailto:sinan@mit.edu}{sinan@mit.edu})}}

\date{} 

\begin{document}
\maketitle
\footnotetext[1]{MIT Institute for Data, Systems, and Society, Cambridge, MA 02142}
\footnotetext[2]{MIT Sloan School of Management, Cambridge, MA 02142}

\begin{abstract}
\textbf{Large Language Models (LLMs) increasingly power generative search engines which, in turn, drive human information seeking and decision making at scale. The extent to which humans trust generative artificial intelligence (GenAI) can therefore influence what we buy, how we vote and our health. Unfortunately, generative search often hallucinates incorrect\cite{liu2023evaluatingverifiabilitygenerativesearch, peskoff-stewart-2023-credible} and dangerous\cite{grant2024} information,posing risks to human health, democracy, and GenAI development. While prior research investigates the reliability, credibility and diversity\cite{liu2023evaluatingverifiabilitygenerativesearch, peskoff-stewart-2023-credible, li2024generativeaisearchengines, sharma2024} of generative search information, no work establishes the causal effect of generative search designs on human trust. Here we execute ~12,000 search queries across seven countries, generating ~80,000 real-time GenAI and traditional search results, to understand the extent of current global exposure to GenAI search. We then use a preregistered, randomized experiment on a large study sample representative of the U.S. population to show that while participants trust GenAI search less than traditional search on average, reference links and citations significantly increase trust in GenAI, even when those links and citations are incorrect or hallucinated. Uncertainty highlighting, which reveals GenAI’s confidence in its own conclusions, makes us less willing to trust and share generative information whether that confidence is high or low. Positive social feedback increases trust in GenAI while negative feedback reduces trust. These results imply that GenAI designs can increase trust in inaccurate and hallucinated information and reduce trust when GenAI’s certainty is made explicit. Trust in GenAI varies by topic and with users’ demographics, education, industry employment and GenAI experience, revealing which sub-populations are most vulnerable to GenAI misrepresentations. Trust, in turn, predicts behavior, as those who trust GenAI more click more and spend less time evaluating GenAI search results. These findings suggest directions for GenAI design to safely and productively address the AI “trust gap.”}    
\end{abstract}

Trust is fundamental to human belief systems and decision-making, influencing the extent to which people rely on AI-generated information and recommendations. As GenAI becomes more prevalent, assessing human trust in GenAI becomes even more critical to understanding how it will shape our world in several ways. First, trust affects the adoption of AI. If users do not trust AI-generated content, they are less likely to use these systems, regardless of their potential benefits\cite{hoff2015}. Second, trust impacts the quality of our decision-making. When we trust AI, we are more likely to integrate its insights into our decisions, potentially leading to better\cite{glikson2020} or significantly worse\cite{vaccaro2024combinations, passi2022overreliance} outcomes. Misplaced trust can lead to over-reliance on AI, which can create critical errors if the AI provides incorrect or biased information\cite{passi2022overreliance, miller2019explanation}. As AI is increasingly deployed in critical domains like healthcare, finance, transportation, and elections, ensuring these systems are trustworthy is essential. Third, AI uses human-in-the-loop feedback to refine algorithms and make them more reliable\cite{amershi2014power}. The cycle of trust and feedback enhances AI performance and drives AI research, fostering innovations that can improve human-AI interactions. If users distrust AI, the quantity and quality of their feedback will decline, hindering safe, effective AI development. 

Studies show that GenAI can produce credible news articles\cite{kreps2022all}, compelling misinformation\cite{spitale2023ai}, clearer and more engaging factual information\cite{huschens2023you}, and political messages that are as persuasive as human-written messages\cite{bai2023artificial, goldstein2024persuasive}. Engineers are currently developing more trustworthy AI through enhanced verifiability, transparency, explainability, and other innovations designed to improve people’s trust in AI\cite{glikson2020, liu2023trustworthy, rawte2023survey, huang2023citation}. We build on this work to examine the conditions under which humans trust GenAI in the context of generative search, estimating the design and contextual factors that influence trust. 

Google, the world’s most popular search engine, handles 8.5B search queries every day, with each of us turning to web search for information three to four times a day on average\cite{Prater2024}. At the same time, GenAI and specifically LLMs are experiencing rapid growth. While a 2023 survey of global workers found that 61\% use or plan to use GenAI\cite{Salesforce2023}, in our 2024 survey 85\% of our experimental subjects reported using GenAI, including ChatGPT and Google’s LLMs, and 63\% of those who reported using GenAI reported using it for information search—the most popular use case in our study (see Table S1 and S2 in the SOM). Unlike conventional web search, which often involves sifting through multiple websites, GenAI can make web search more convenient and efficient by providing direct responses to information queries with details summarized from different sources. However, the stochastic nature of LLM outputs can produce inaccurate and unsupported information, commonly known as “hallucinations”\cite{liu2023evaluatingverifiabilitygenerativesearch, grant2024, liu2023trustworthy, rawte2023survey, huang2023citation, memon2024search, zhang2023siren}. Given our reliance on web search for information and the potentially dramatic influence it can have on our decision making, it is important, as generative search is widely adopted, to understand how the generative search experience differs from traditional search and how people perceive and potentially rely on generative search information differently.

Although generative search is one of the most popular and fastest growing GenAI use cases, scant research addresses this application. Audit studies have assessed the quality of information and citations returned in generative search and noted its proclivity to produce errors\cite{liu2023evaluatingverifiabilitygenerativesearch,li2024generativeaisearchengines,memon2024search}. Other studies compared generative and conventional search through randomized experiments, examining search performance and user experiences with ChatGPT and Google search\cite{xu2023chatgpt}, comparing product search behaviors with generative and conventional search\cite{spatharioti2023comparing}, and assessing selective exposure in generative search\cite{sharma2024}. However, because these experiments randomly assign people to different models, it is difficult to assess whether AI or differences in the information provided by different models drive the results. These experiments do not explain how generative search design or heterogeneity in human factors, like participants’ education, employment, or experience with GenAI, affect users’ trust.

To first understand the extent of global exposure to GenAI search results, we executed 11,372 queries, using serpAPI’s Google Search Engine Results API to generate 79,604 real world GenAI and traditional search results across seven countries by setting the country location of each API call. The search queries were randomly sampled from nine data sources containing real user searches including “Google Natural Questions,” “Most-Searched Google Queries,” “Covid-Related Search Queries,” and “Amazon Shopping Queries” (a full list of query datasets is provided in the SOM). We used GPT-4o-mini to label the style of each query (whether it was a question, a statement or navigational) and to generate one topic label per query for a random subset of queries. We then built a topic list from these initial LLM labels and prompted gpt-4o-mini to label the rest by picking one topic per query from this list that best fit the query. We then manually grouped LLM-generated topic labels into seven categories: General Knowledge, Health, Internet / Technology / Media, Shopping, Lifestyle, Business / Finance / Employment, and Covid, where Covid is a special health category which includes all queries that mention “Covid” or “Coronavirus.” 

\begin{figure}[hb!]
    \centering
    \includegraphics[width=1\linewidth]{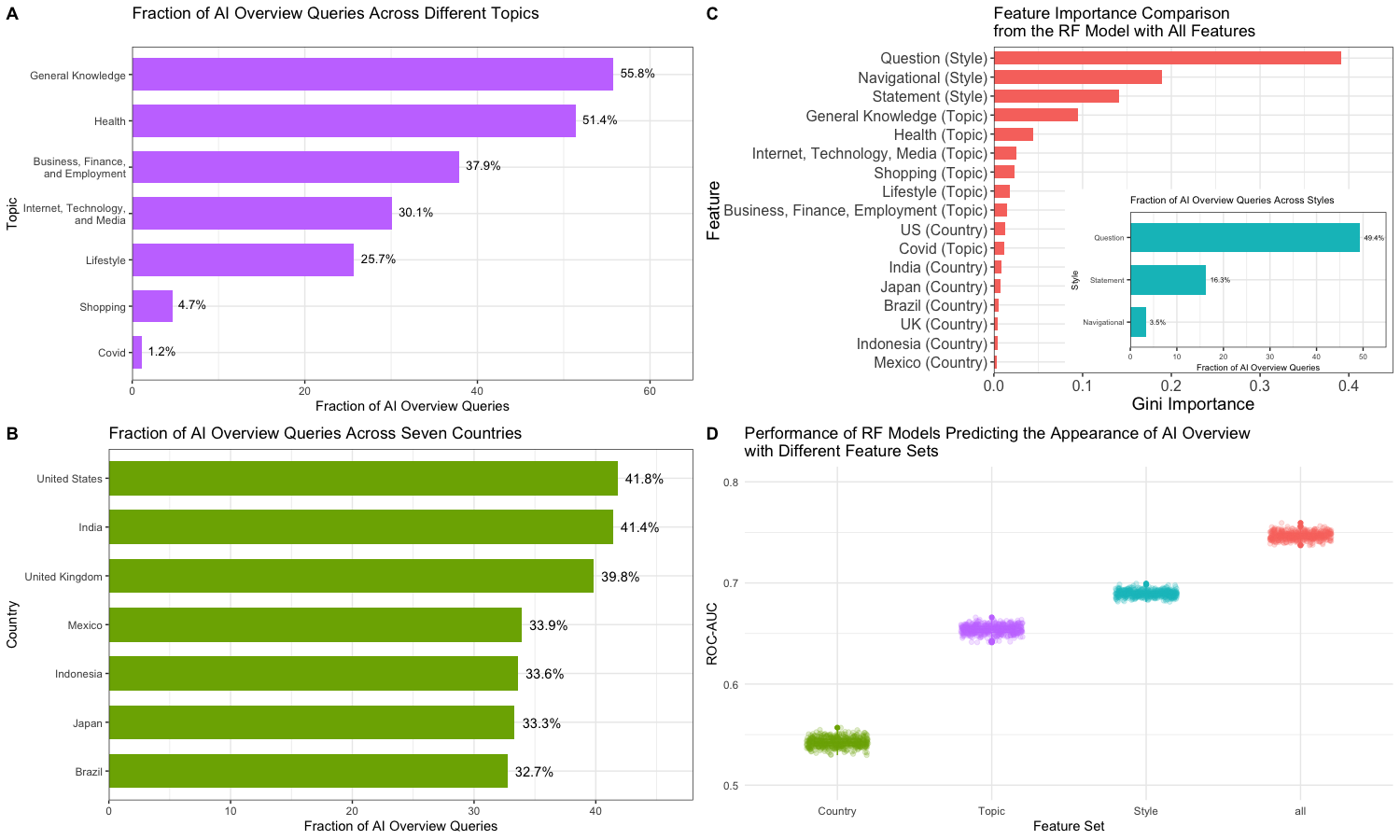}
    \caption{\footnotesize \textbf{Global Exposure to GenAI Search Results.} Figure 1 displays (A) the fractions of query searches with AI results across seven topics. Queries mentioning “Covid” or “Coronavirus” were removed from “Health” and grouped together under the “Covid” group; (B) the fractions of queries with AI results across the seven countries that had the AI Overview feature publicly available at the time of our data collection: US, UK, India, Japan, Indonesia, Mexico, and Brazil; (C) feature importance from a random forest model using all features to predict whether a query search returns AI results. Features (country, style, topic) are ranked by Gini importance, with higher scores indicating greater importance of the features; (C inset) the fraction of queries with AI results, searched in all seven countries, across three query styles—question, statement and navigational; and (D) a comparison of the performance of four random forest models in predicting whether a search would have an AI overview. Each model was trained with a different feature set (country, topic, style, or all) using repeated 5-fold cross-validation, generating 500 AUC scores per model. The figure presents a box plot overlaid on a scatterplot of the 500 AUC scores for each model.}
    \label{fig:fig1}
\end{figure}

\vspace{1em}
The data show GenAI search results are globally pervasive but vary greatly by topic and query style. Over half of all Health (51\%) and General Knowledge (56\%) queries returned GenAI results while only 5\% of Shopping queries and 1\% of Covid queries returned GenAI results (Fig. \ref{fig:fig1}A). GenAI appeared with similar frequencies across countries with queries executed in the U.S. returning GenAI results 42\% of the time and in Brazil 33\% of the time (Fig. \ref{fig:fig1}B). The style of the query was extremely influential in determining whether users saw GenAI results, with questions returning GenAI answers 49\% of the time, statements 16\% of the time and navigational searches returning GenAI only 4\% of the time (Fig. \ref{fig:fig1}C, inset). Random forest models indicate that the style of queries explained the most variance in whether searches returned GenAI results followed closely by topics, while the country of origin explained very little of the variation (Fig. \ref{fig:fig1}C, D).

To evaluate people’s trust in GenAI as an information search source and to identify the design factors influencing trust, we conducted a large-scale, pre-registered, randomized experiment involving a sample of 4,927 participants representative of the adult population of the United States. We randomly assigned participants to GenAI and traditional search conditions across nine pre-defined search queries, covering topics from gun control to adult vaccine schedules, and presented them with generative responses directly collected from Google’s AI Overviews. Participants in both groups saw identical information presented as either generative search or traditional search results and were informed that the information was either created by GenAI or not. They were then asked to report how much they trusted the information in GenAI or traditional search using a 5-item message credibility measure that has been validated extensively in prior work\cite{flanagin2007role}. 

Our trust measure mirrors work that quantifies human perceptions of web information. Early research that measured trust in search results and websites examined the message, source, sponsor, and site\cite{flanagin2007role, appelman2016measuring, metzger2003credibility}. In this literature, trust is commonly conceptualized across the five dimensions of accuracy, believability, consistency, completeness, and trustworthiness, all measured on a Likert scale\cite{flanagin2007role, appelman2016measuring, sundar1999exploring}. We elicited participants’ trust in generative and traditional search, across these five trust elements, by asking them to report their agreement with the following statements on a 7-point Likert scale, from strongly agree to strongly disagree: “I think this response is accurate/believable/biased/complete/trustworthy”. We then assessed the reliability and internal consistency of our 5-item trust rating using Cronbach’s alpha, which measures the shared variance among the items relative to their overall variance\cite{flanagin2007role, appelman2016measuring, cronbach1951coefficient}. Cronbach alpha values were all greater than 0.87, suggesting high reliability and internal consistency in our trust measure (see Table S3 in the SOM). To verify its construct validity, we conducted a confirmatory factor analysis (CFA), in which the factor loadings were all positive, significant and exceeded the recommended cutoff of 0.4 across all search tasks\cite{gefen2000structural}, indicating high convergent validity and a strong relationship between the five observed items and a latent trust factor (see Table S4 in the SOM). 

We also asked respondents to report their willingness to “share this response with a friend who is interested in this question/topic,” again on a 7-point Likert scale. The willingness to share measures an actionable intention to provide information to others and has been widely used in previous work examining people’s perceptions of and belief in online information\cite{pennycook2021practical}. Substantial evidence suggests a strong correspondence between self-reported willingness to share and actual levels of sharing of online content\cite{mosleh2020self}. We consider the intension to share GenAI results a proactive indicator of confidence. While the willingness to share likely measures both trust and participants’ perceptions of the value of search information for others, people are more willing to share responses they find credible and the perceived value to others is held constant in our study as control and experimental groups saw identical information presented as either GenAI or traditional search. Sharing intensions are also important as they indicate the extent to which GenAI may diffuse from person to person in human information seeking and problem-solving tasks. 

In addition to comparing GenAI to traditional search, we examined the impact of four randomized GenAI designs—the use of references, uncertainty highlighting, explanations, and social feedback—on trust in generative AI. These trust cues mirror those studied in previous work on peoples’ trust of web-based information\cite{glikson2020,wathen2002believe,schwarz2011augmenting,fogg2001makes} and represent GenAI search designs prevalent in today’s market. We designed these trust cues to be orthogonal to the generative information itself to mitigate the influence of model quality on our results and to enhance the robustness and generalizability of our findings. We also presented different versions of three of the randomized GenAI designs—references, uncertainty highlighting, and social feedback—to understand how variations in these designs affect trust in within-subjects experiments. 

First, the use of references and reference links is a major design challenge for GenAI. Although providing supporting information can enable trust, GenAI frequently hallucinates inaccurate and unsupported references and reference links, creating a breeding ground for misinformation and dangerous guidance on health related and other behaviors. We randomly assigned some participants to receive generative search results with references and reference links and compared them to a control group that saw generative information without references. Among those who were assigned to the reference link condition, we randomized whether the search responses cited were completely valid and accurate references or contained hallucinated, broken or invalid references. 

Second, uncertainty highlighting is a common generative AI design that highlights text in which models have high or low confidence in different colors. Uncertainty highlighting reveals when GenAI is confident in the information it provides and when it is not. Designers expect high confidence highlighting to increase trust in GenAI information and low confidence highlighting to do the opposite. To test the effect of both high and low confidence uncertainty highlighting, we randomized the generative responses that users in the uncertainty highlighting condition saw to either display strictly low certainty (orange) highlighting or both high (green) and low (orange) certainty highlighting. 

Third, generative search engines frequently display social feedback that shows users the fraction of other users who have found the generative information helpful, a common design in social information channels. Prior work has shown that social feedback can influence trust in and use of Covid-19 vaccines and that when greater proportions of their peers trust vaccines, people are more willing to accept vaccines themselves\cite{moehring2023providing, holtz2020interdependence}. To test the effect of both positive and negative social feedback, we randomized whether generative search responses where displayed with a randomly selected fraction of “users who found this helpful” from either a uniform distribution between 65\% and 95\% in the positive social feedback condition, or a uniform distribution between 5\% and 35\% in the negative social feedback condition, and compared them to each other and to a condition without social feedback.

Fourth, generative search engines typically include explanations of how GenAI models work. To test the explanation design, we presented the explanation displayed by Google AI Overviews to participants randomized into the explanations condition. This text explains how GenAI search works, describes the “pros” and “cons” of generative search, and explicitly describes how generative search uses “high quality web results” to “significantly mitigate some of the known limitations of LLMs, like hallucination or inaccuracies” (the exact text is shown in Section 1.5 of the SOM). Designers expect explanations to increase trust in generative search outputs and, as described, to mitigate against limitations of LLMs that may reduce trust. 

Finally, a large literature demonstrates heterogeneity in people’s perception of and interaction with AI based on their age, education, jobs, and prior knowledge and experience with the technology\cite{sharma2024, kreps2022all, mahmud2022influences,araujo2020ai}. We explored the heterogeneity of people’s trust by examining how people with different demographics, educational and professional backgrounds, political leanings, and levels of familiarity with AI, trust generative information differently.

We evaluated GenAI trust effects across nine search queries representing five policy debates and four topics from Pew Research Center’s list of “Top US problems”\cite{PewResearch2023} to understand how participants’ trust in GenAI varies by topic and information seeking task. The five policy debate topics represent information seeking and opinion formation related to current affairs relevant to the adult US population, adapted from \cite{stagnaro2023no}. They include the following queries: Will increased gun control lead to more or less crime? Is implementing universal healthcare in the US negative or positive for the health of citizens? Does more immigration take jobs from or create jobs for US citizens? Does increased civilian oversight of police lead to more or less crime? Will raising taxes on wealthy individuals within the US increase or decrease tax revenue? We then adapted Pew Research Center’s list of “Top US problems” into search questions by inputting them into Google search and utilizing the top suggested search term returned by Google for each topic, which produced the following search topics: “government policy to fight inflation,” “jobs replaced by AI,” “how often do adults need vaccines,” and “harmful effects of climate change.”

We analyzed average, between-subjects, treatment effects using the following linear model:
\begin{equation}
Y_{iq}=\alpha_0 + \sum_t\beta_t(\text{Treatment}_{it}) + \gamma X_i + \eta_q + \epsilon_{iq}
\label{spec:btw}
\end{equation}
in which $Y_{iq}$ represents our trust metric or the willingness to share of person $i$ on search task $q$; $\alpha_0$ is a constant term; $t$ indexes the Treatment, for example reference links or uncertainty highlighting; $X_i$ is a vector of pre-treatment, individual-level covariates including age, gender, ethnicity, education level, career industry, political leaning, and the frequency of GenAI and generative search use; $\beta_t$ represents the average treatment effects of GenAI and different GenAI designs $t$; $\eta_q$  are search task fixed effects; and $epsilon_iq$ are heteroskedasticity-robust standard errors. We also measured conditional average treatment effects for subpopulations using this estimating equation on subsamples of participants, for example with higher or lower levels of education or greater or less familiarity with GenAI and for different search tasks, for example gun control or climate change.

We used a two-way fixed effects model to analyze the within-subjects treatment effects of different versions of the references, uncertainty highlighting, and social feedback designs as follows:
\begin{equation}
Y_{iq}=\beta(\text{Variation}_{iq}) + \lambda_i + \eta_q + \epsilon_{iq}
\label{spec:within}
\end{equation}
where $Y_{iq}$ represents trust or the willingness to share; $\text{Variation}_{iq}$  represents the variation of the baseline treatment, for example valid versus hallucinated reference links or positive versus negative social feedback; $\lambda_i$ is an individual level fixed effect; $\eta_q$  are search task fixed effects; and $\epsilon_{iq}$ are standard errors clustered at the individual and search task levels. We also measured conditional average treatment effects for different search tasks, for example estimating the effects of positive versus negative social feedback on trust when participants searched for information about gun control versus climate change. 

\begin{figure}[t!]
    \centering
    \includegraphics[width=1\linewidth]{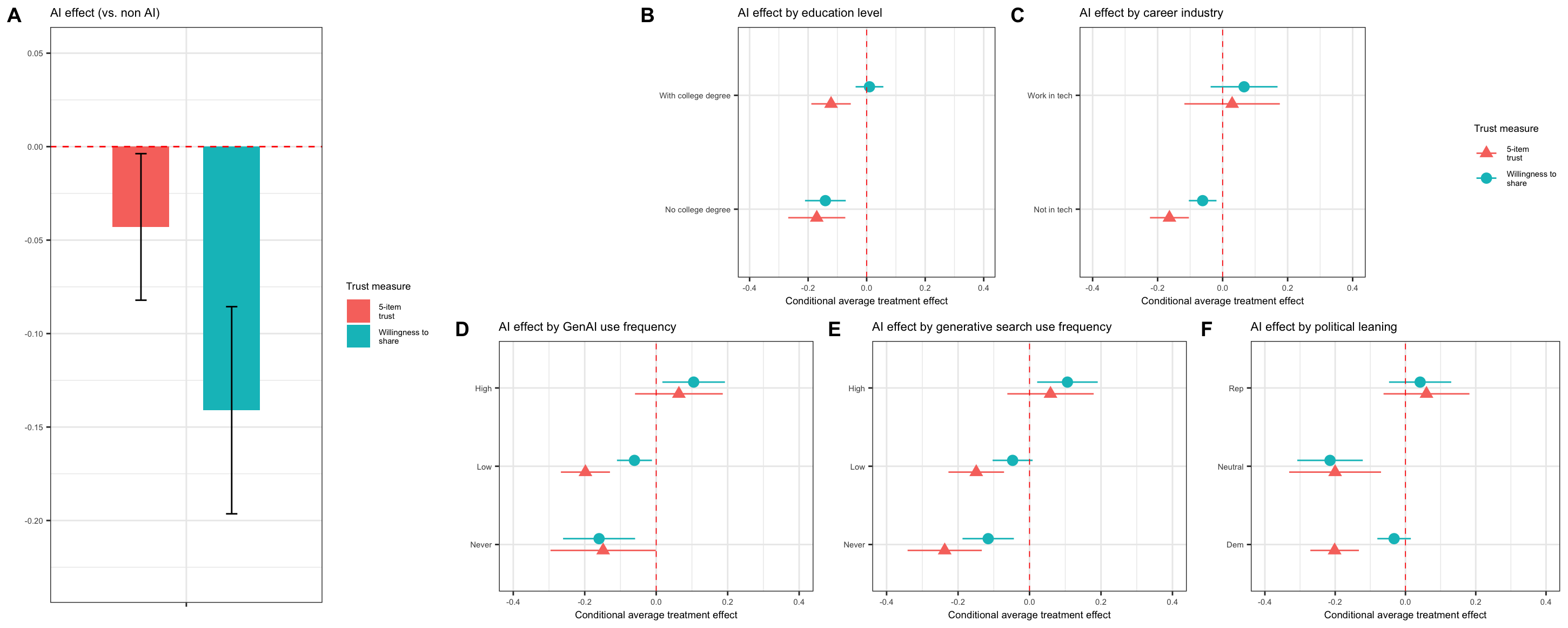}
    \caption{\footnotesize \textbf{Generative Search and Trust.} Figure 2 displays (A) the average treatment effects of providing
generative search information compared to the traditional search levels of trust as well as the heterogeneity of
these treatment effects by subjects' (B) education level, (C) employment in technology related industries,
frequency of (D) GenAI and (E) generative search use, and (F) political leaning.}
    \label{fig:fig2}
\end{figure}

The experimental results indicate that, on average, generative search significantly reduces participants’ trust in and willingness to share search results compared to traditional search (Fig. \ref{fig:fig2}A). As the information provided in the GenAI and traditional search conditions is identical, these results suggest that participants’ belief that AI is producing the results causes them to trust the information less and to be less willing to share it. However, different subpopulations react differently to generative information. Participants with higher levels of education (Fig. \ref{fig:fig2}B), operationalized as having a college degree or higher, trust GenAI information more and are willing to share it significantly more than those with no college degree. Participants with work experience in technology related industries (Fig. \ref{fig:fig2}C) trust GenAI information significantly more and are willing to share it significantly more than those with no experience working in technology industries (Fig. \ref{fig:fig2}D). Those who use GenAI (Fig. \ref{fig:fig2}E) and generative search (Fig. \ref{fig:fig2}F) frequently trust generative search results significantly more and are significantly more willing to share them than those who use these technologies less or have never used them. Finally, Republicans trust generative search results significantly more than Democrats and those who report having neutral political leanings (Fig. \ref{fig:fig2}G).

\begin{figure}[h!]
    \centering
    \includegraphics[width=1\linewidth]{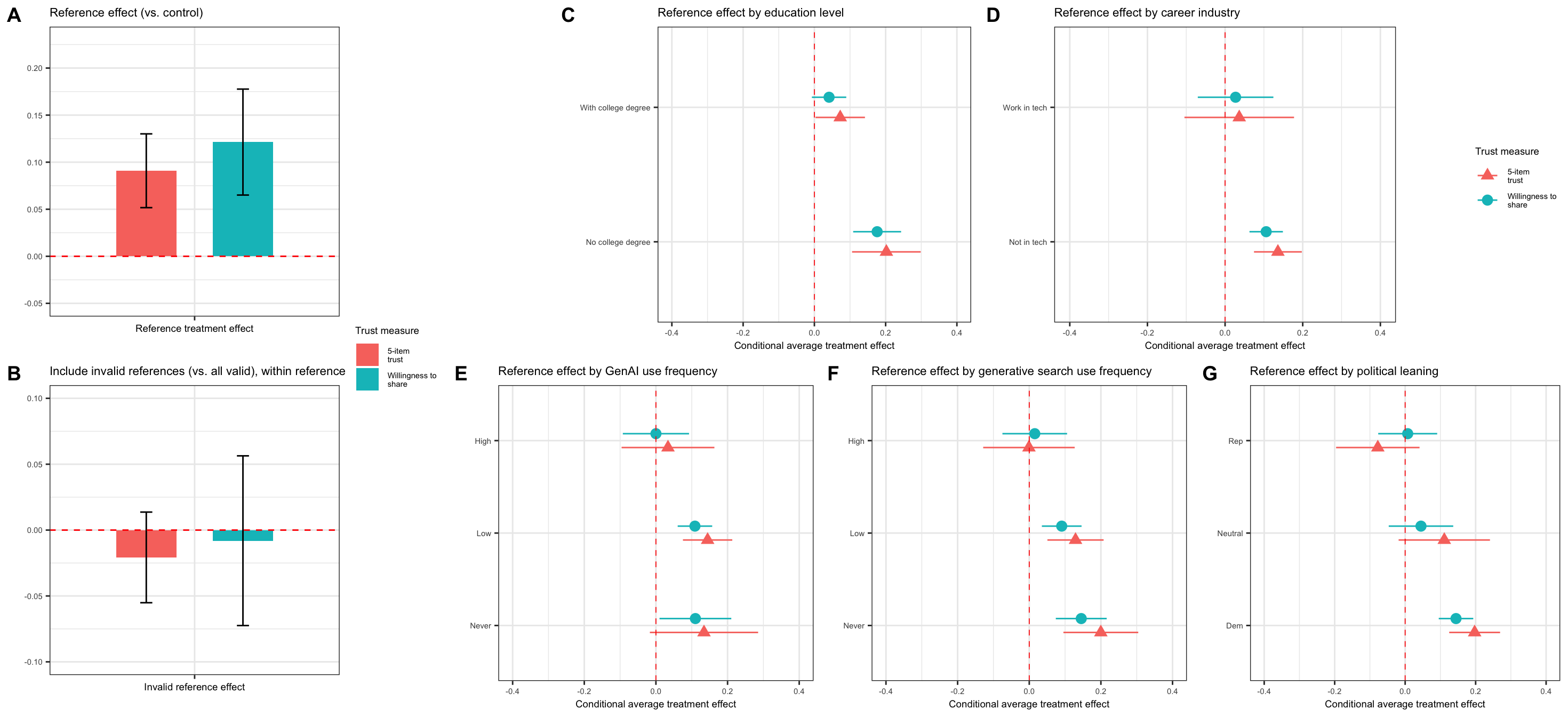}
    \caption{\footnotesize \textbf{Trust Effects of Valid and Invalid GenAI References.} Figure 3 displays (A) the average treatment effects of providing references and reference links in support of the information provided in GenAI search results on participants’ trust and willingness to share GenAI search results. Search task responses were randomized to contain either valid and correct or invalid and hallucinated references. Panel (B) displays differences in the effects of providing either valid or invalid GenAI references and links. The heterogeneity in the average treatment effects of providing references links, pooled across both valid and invalid links, across subjects (B) education level, (C) employment in technology related industries, frequency of (D) GenAI and (E) generative search use, and (F) political leaning are displayed in the remaining panels.}
    \label{fig:fig3}
\end{figure}

Providing reference links significantly increases participants’ trust in and willingness to share generative search results (Fig. \ref{fig:fig3}A). Sadly, the increases in trust and willingness to share caused by the provision of references are unaffected by whether the references and links are valid or invalid (Fig. \ref{fig:fig3}B). References and reference links increase trust and the willingness to share generative search results significantly more for participants who have lower education levels (Fig. \ref{fig:fig3}C) and for participants who do not work in technology related industries (Fig. \ref{fig:fig3}D). The frequency of GenAI (Fig. \ref{fig:fig3}E) and generative search (Fig. \ref{fig:fig3}F) use does not significantly affect the degree to which references impact trust in or willingness to share GenAI search results. But participants’ political leaning does have an effect. Democrats trust GenAI outputs with references and reference links significantly more than Republicans and are significantly more willing to share them (Fig. \ref{fig:fig3}G). 

\begin{figure}[t!]
    \centering
    \includegraphics[width=1\linewidth]{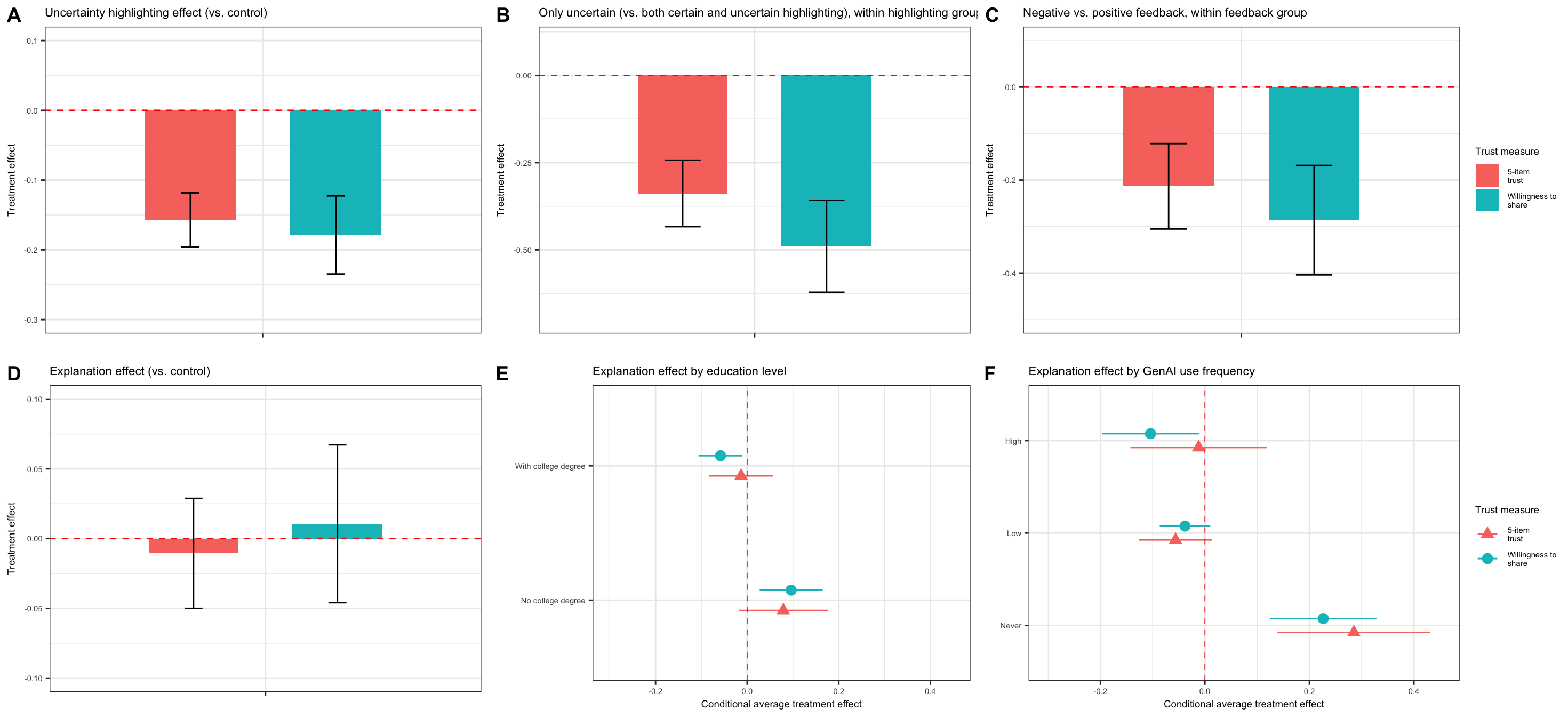}
    \caption{\footnotesize \textbf{Trust Effects of Uncertainty Highlighting, Social Feedback and GenAI Explanations.} Figure 4 displays (A) the average treatment effects of uncertainty highlighting and (B) the effects of low certainty highlighting compared to both high and low certainty highlighting, as well as (C) the effects of negative versus positive social feedback and (D) an explanation of how GenAI works to create generative search results on participants trust in and willingness to share generative search information. Panel (E) displays variation in the treatment effects of providing explanations by participants education levels and Panel (F) displays variation in the treatment effects of providing explanations by participants prior frequency of GenAI use.}
    \label{fig:fig4}
\end{figure}

Uncertainty highlighting uniformly reduces trust in and willingness to share GenAI outputs (Fig. \ref{fig:fig4}A). This is true when only low certainty highlighting is displayed and when high and low certainty highlighting is displayed, although the reductions in trust and willingness to share are greater when only low certainty highlighting is displayed, compared to a condition that highlighted text snippets in which the model had high and low levels of certainty (Fig. \ref{fig:fig4}B). Negative social feedback significantly reduces trust in and willingness to share GenAI outputs compared to positive social feedback (Fig. \ref{fig:fig4}C). Providing explanations has no effect on trust in and willingness to share GenAI outputs, on average (Fig. \ref{fig:fig4}D). However, participants with lower education levels experience increases in trust and significant increases in willingness to share GenAI outputs when they are provided with explanations of how GenAI models work to create generative search results (Fig. \ref{fig:fig4}E). Participants who report never having used GenAI experience significant increases in trust and willingness to share GenAI outputs when they are provided with these explanations (Fig. \ref{fig:fig4}F).

\begin{figure}[b!]
    \centering
    \includegraphics[width=1\linewidth]{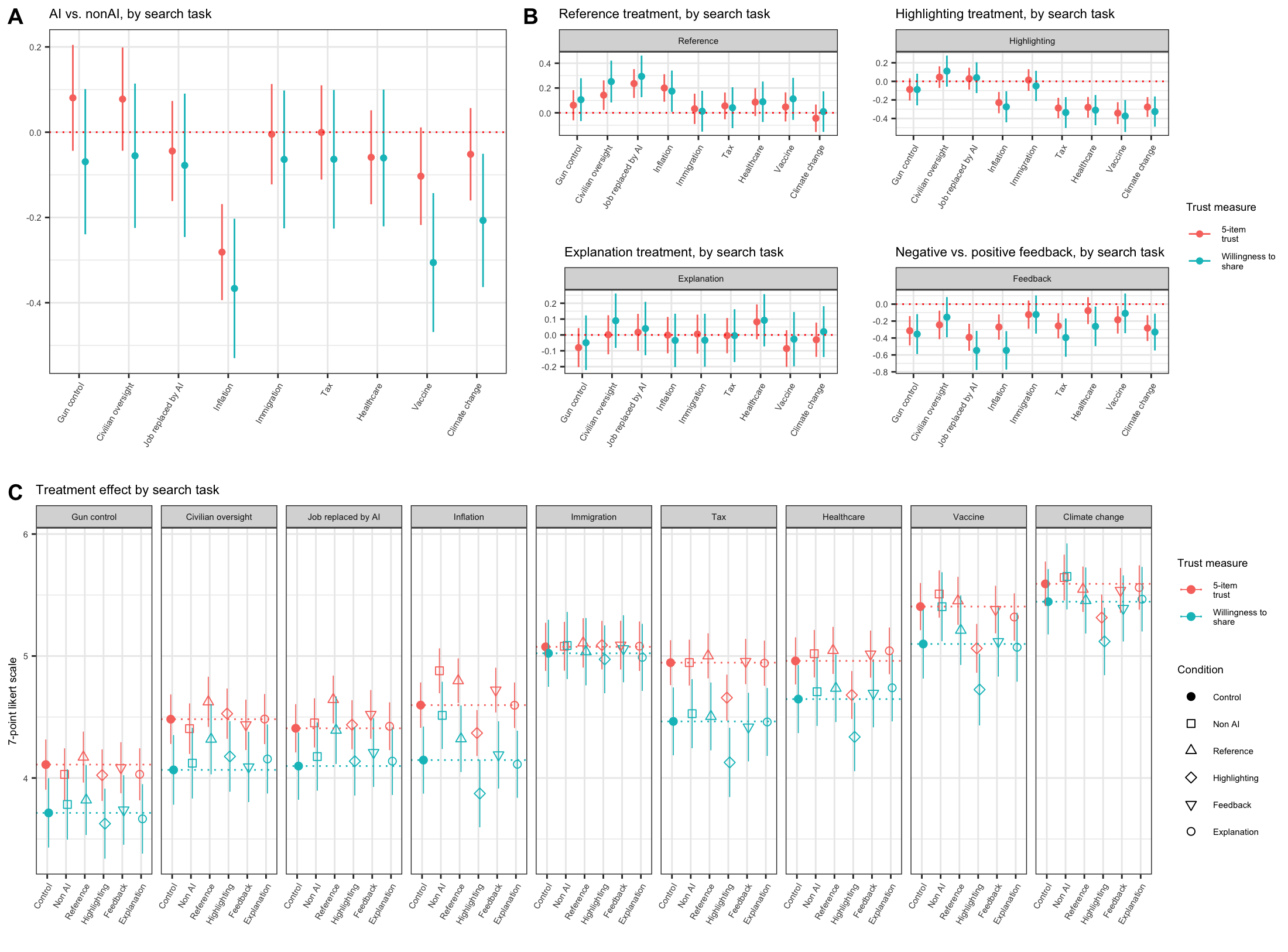}
    \caption{\footnotesize \textbf{Trust Effects of GenAI Designs Across Search Topics.} Figure 5 displays (A) the average treatment effects of providing generative search information on trust in and willingness to share GenAI information, compared to the traditional search levels of trust, across the nine search topics in our study. Panel (B) displays the effects of GenAI designs that include references, uncertainty highlighting, explanations and negative versus positive feedback on trust in and willingness to share GenAI information, across the same nine search topics. Panel (C) displays levels of trust across traditional search and all the GenAI search designs, across the nine search topics.}
    \label{fig:fig5}
\end{figure}

Trust and willingness to share generative information also vary by search topic. As shown in Fig. \ref{fig:fig5}A, generative search reduces participants’ trust in and willingness to share search results when they are searching for information related to inflation, compared to traditional search. Generative search also significantly reduces participants’ willingness to share search results related to adult vaccine schedules and climate change, compared to traditional search, although the corresponding reductions in trust are not significant. Design features also display heterogeneous treatment effects across topics. The provision of reference links significantly increases participants’ trust in and willingness to share GenAI search results when they are searching for information on whether jobs will be replaced by AI, civilian oversight of police and government policies to fight inflation, but not when they are seeking information about gun control, immigration, tax policy, healthcare, vaccine schedules or the harmful effects of climate change (Fig. \ref{fig:fig5}B). Uncertainty highlighting significantly reduces participants’ trust in and willingness to share GenAI results when they are searching for information about inflation, tax policy, healthcare, vaccines and climate change, but not when they are seeking information about gun control, civilian oversight of police, or whether jobs will be replaced by AI or substituted by immigration (Fig. \ref{fig:fig5}B). Negative social feedback significantly reduces trust in GenAI for all topics except immigration and healthcare and significantly reduces participants’ willingness to share GenAI output for all topics except immigration, vaccines, and civilian oversight of police (Fig. \ref{fig:fig5}B). Explanations have no effect on trust or the willingness to share information on any topic.

\begin{figure}[b!]
    \centering
    \includegraphics[width=1\linewidth]{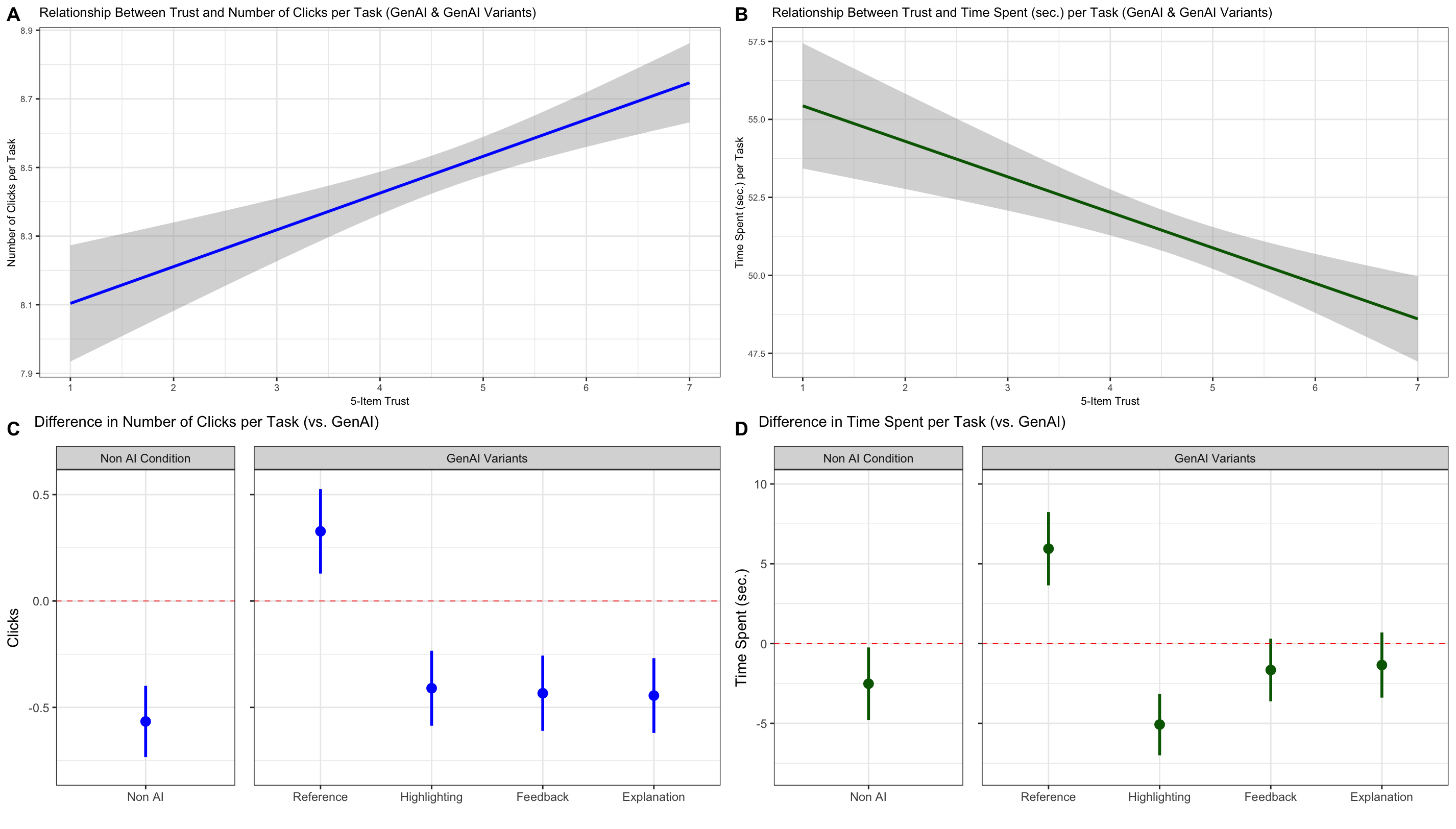}
    \caption{\footnotesize \textbf{Effects of GenAI and GenAI Trust on Clicks and Time Spent.} Figure 6 displays (A) results of regression analysis of the 5-item trust score on the number of clicks per task and (B) the 5-item trust score on time spent in seconds per task, using all ratings from participants in the GenAI and the GenAI variant conditions. Panel (C) shows the difference in the number of clicks per task of the traditional search condition, the reference group condition, the uncertainty highlighting condition, the social feedback condition and the AI explanation condition, relative to the GenAI condition. Panel (D) shows the difference in the time spent per task in seconds of each condition relative to the GenAI condition.}
    \label{fig:fig6}
\end{figure}

Trust in search information, across traditional search and all the GenAI search designs, exhibits significant variation across topics as well, with participants reporting the highest levels of trust and willingness to share search information for information on climate change and the lowest levels of trust and willingness to share information on gun control across all conditions (Fig. \ref{fig:fig5}C). Interestingly, participants’ trust in and willingness to share search information is highly correlated for every topic except taxes and inflation. This result suggests that participants trust search information on taxes and inflation more than they are interested in sharing this information with others.

Trust also predicts behavior. Participants who report greater trust in GenAI click more on GenAI search results (Fig. \ref{fig:fig6}A) but spend less time evaluating and interacting with GenAI search content (Fig. \ref{fig:fig6}B). These behavioral differences suggest that trust in GenAI inspires search engine users to click on GenAI results more quickly and perhaps, given the less time they spend interacting with the content, with less discernment or critical evaluation. Compared to the GenAI condition, participants randomly assigned to traditional search click less on (Fig. \ref{fig:fig6}C) and spend less time evaluating (Fig. \ref{fig:fig6}D) search results, which reinforces the behavioral results from Figs. 6A and 6B. Compared to the GenAI condition, participants assigned to the reference AI condition click more on (Fig. \ref{fig:fig6}C) and spend more time evaluating (Fig. \ref{fig:fig6}D) search results. Compared to the GenAI condition, participants assigned to the uncertainty highlighting, social feedback and AI explanations conditions click less on search results (Fig. \ref{fig:fig6}C), and participants assigned to the uncertainty highlighting condition spend less time evaluating search results (Fig. \ref{fig:fig6}C). In our experiment, reference links encourage clicking and time spent, while uncertainty highlighting, social feedback and AI explanations all reduce clicking, and only uncertainty highlighting reduces time spent. The data suggest that trust in AI mediates our reliance on and critical discernment of AI search results. They also suggest that GenAI designs affect clicking and the time spent evaluating search result content.

We conducted multiple checks during data collection and analysis to ensure the robustness of our results. First, average attrition rates after treatment assignment were low at 2.62\% and results of a logistic regression predicting attrition by treatment assignment showed no significant differences between treatment groups, lending credibility to the robustness of our treatment assignment and suggesting attrition did not affect our results (see Table S7 in the SOM). Second, randomization checks assessing the balance of all individual level covariates across treatment groups found no significant variation in 47 of 51 covariates. The four covariates that showed significant differences were due, in large part, to the small sample sizes in their respective categories combined with random chance (see Table S8 in the SOM). Third, we performed randomization inference for between-subjects and within-subjects treatment effects, testing the sharp null of no individual-level treatment effects, implemented through randomly redistributed treatment and treatment variation assignments and found that except for the effect of traditional search on our five item trust measure, the significance levels of all of our results were unchanged (see Tables S21 and S22 and the accompanying text in the SOM). Fourth, we implemented three Placebo tests. Although the GenAI, traditional search, and explanation groups did not have any variations, the treatment variation assignment algorithm was still implemented in these groups. As participants were randomly assigned to identical baseline and variation treatments in these groups, we should observe no differences in their effects. As expected, none of the three groups showed significant differences, lending credibility to the randomization and to the treatment effects we observe in the other groups (see Table S23 in the SOM).

Although we conducted a large, representative experiment with many robustness checks, our work is not without its limitations. First, as with any lab experiment, results may vary in the field. Unfortunately, only search engines themselves can conduct GenAI design experiments in the wild and they are unlikely to report such results publicly, making lab experiments one of the only public mechanisms with which to test GenAI designs. To mitigate the drawbacks of lab experimentation, we made our experiment as realistic as possible, for example providing clickable reference links in a realistic interface that mimicked search designs and including randomly selected fractions of positive and negative social feedback. We encourage GenAI designers to report the results of GenAI field experiments publicly to understand how they may differ from those found in the lab. Second, we explicitly chose consequential search topics on public affairs and “top problems” likely to sway meaningful human decision making, like how people vote. However, generative search is also used for more mundane tasks like product search or DIY solutions to common household problems. Results may vary for such search topics, and we encourage future research on other types of search tasks to understand how generative search may be best applied. Third, we focused exclusively on the U.S. market, even though GenAI is being deployed globally. While our approach produced precise estimates for a study sample representative of the adult U.S. population, trust in GenAI may vary significantly across cultures and languages. We hope future work will reveal these differences and how human trust in GenAI varies across geographies. Finally, GenAI design is rapidly evolving. It is therefore unlikely that any AI design experiment will be definitive. But we believe experimental results of current AI designs reveal more fundamental aspects of human trust, which in turn should inform the evolution of AI design. Given the crucial importance of trust in the use of AI, experimental evaluations of current AI systems will help us design the safest and most effective AI over time.

Despite these limitations, ours is the first experiment testing how GenAI search designs affect trust, holding AI models and their outputs constant. Our experiment shows that although participants trust GenAI less than traditional search on average, Republicans and people with higher levels of education, and experience with GenAI tend to trust it more. Our results also show which designs induce trust, even when trust is unwarranted, and which populations are most vulnerable to such AI misrepresentations. Perhaps most striking, references and reference links induce trust, even when they are hallucinated, invalid or broken. In other words, the veneer of rigor in GenAI design creates trust even when references and links are not rigorous. Furthermore, Democrats and those with lower levels of education and less experience with GenAI are induced to trust GenAI outputs more when valid or invalid references are provided than Republicans or those with higher levels of education and experience with GenAI, revealing who is specifically vulnerable to this veneer of rigor. Our results suggest that GenAI interface designs can create trust, holding the trustworthiness of the models constant. They also suggest that trust and GenAI design mediate clicking and the amount of time we spend evaluating GenAI search result content.  Trust in AI, and reliance on information produced by AI, is therefore not only a function of the trustworthiness and reliability of the underlying models, but also of how human-AI interfaces are designed. This realization should elevate AI interface design to the level of foundation model reliability in their relative importance in the AI trust debate.

\clearpage
\bibliographystyle{unsrt}
\bibliography{mainbib}

\clearpage
\section*{Acknowledgments}
We thank members of the MIT Initiative on the Digital Economy and the MIT Social Analytics Lab for their helpful comments on earlier drafts of this manuscript. The study was approved by the MIT Institutional Review Board (IRB Protocol E-5570). Funding: The authors thank the MIT Initiative on the Digital Economy and its founding sponsor, Accenture, for financial support of this research. Author contributions: Conceptualization: H.L and S.A.; Methodology: H.L and S.A.; Data curation, software and formal analysis: H.L.; Investigation: H.L and S.A.; Writing: H.L and S.A.; Visualization: H.L and S.A.; Project administration and resources: H.L and S.A.; Supervision: H.L and S.A.; Validation: H.L. Competing interests: The authors declare no competing interests. Data and materials availability: Preregistration, all original analysis code, aggregated data and annotated code sufficient to replicate the results are available here: \href{https://osf.io/2du7x/}{https://osf.io/2du7x/}.

\renewcommand{\thefigure}{S\arabic{figure}}  
\setcounter{figure}{0}  
\setcounter{equation}{0}
\vspace*{2cm}
\begin{center}
    {\LARGE Human Trust in AI Search:} \\
    \vspace{0.5em}
    {\LARGE A Large-Scale Experiment} \\
    \vspace{1em}
    {\Large Supporting Online Materials} \\
    \vspace{3em}
\end{center}

\tableofcontents
\section{Estimating Global Exposure to GenAI Search}
To understand global exposure to GenAI search, we estimated how frequently GenAI Search responses are provided to users searching Google, and how the frequency varies by search query topic, style, and country. We collected 79,604 Google search results, obtained by running 11,372 real-world user search queries across 7 countries through Google Search. The data was collected between October 8, 2024 and November 8, 2024.  

\subsection{Query Selection}
Our search results consist of 11,372 real-world user search queries, randomly sampled from the following data sources, with all non English queries filtered out: 
\begin{itemize}
    \item 3561 queries from Google’s \href{https://ai.google.com/research/NaturalQuestions}{Natural Questions}.
    \item 3386 queries from Microsoft’s \href{https://microsoft.github.io/msmarco/}{MSMARCO}
    \item 3130 queries from \href{https://quoradata.quora.com/First-Quora-Dataset-Release-Question-Pairs}{Quora Question Pairs}.
    \item 428 queries from 2015-2020 Google Trends keywords (US and Global) 
    \item 283 queries from \href{https://www.semrush.com/blog/most-searched-keywords-google/}{most-searched Google queries (US and Global)}
    \item 200 coding queries from \href{https://github.com/microsoft/Search4Code}{Microsoft’s Search4Code}
    \item 168 queries from a \href{https://figshare.com/articles/dataset/Search_query_lists/12398309}{COVID-related search query dataset}
    \item 116 queries from a \href{https://huggingface.co/datasets/trec-product-search/product-search-2024-queries}{product search query dataset}
    \item 100 queries from \href{https://github.com/amazon-science/esci-data}{Amazon shopping query dataset}
\end{itemize}

\subsection{GenAI Search Result Collection}
To survey whether these queries would induce an GenAI response if people search them on Google, we used \href{https://serpapi.com}{serpApi’s Google Search Engine Results API} to collect Google search results for these 11,372 search queries. If a Google search returned an AI Overview (Google's GenAI Search response), the search result output from the corresponding API call will include an AI Overview block and, for such results, we recorded a GenAI search result. For each unique query, we implemented 7 searches---one in each of the 7 countries where AI Overviews was a publicly available feature of Google Search at the time of our data collection (US, UK, India, Mexico, Brazil, Japan, and Indonesia)---by setting the location of the API call to that country. This generated a total of 79,604 search results ($11372 \times 7$).

\subsection{Query Categorization}
Tables \ref{si-tab:query-style} and \ref{si-tab:query-topic} present the descriptive statistics of the search queries we used, based on their style and topic. We used GPT-4o-mini to label the style (question, statement or navigational) of each query. We used the following prompt: “Identify the format of the following search engine query: whether it is a statement, an interrogative query (question), or a navigational query (an internet search with the clear intent of finding a specific website or web page; a navigational query can usually be the name of a brand, platform, organization, or specific URL, intending to navigate directly to that destination).”

For topic labeling, we first used GPT-4o-mini to generate one topic label per query for a random subset of queries, using the following prompt: “Label the following search engine query with one topic based on the subject matter.” We built a topic list from these initial LLM labels, and prompted gpt-4o-mini to label the rest by picking one topic per query from this list. We used the following prompt: “Label the following search engine query with one topic based on the subject matter. Your answer should be a topic label from the provided list. If the query doesn't fit in to any of the given topics, come up with your own topic.” Finally, we manually grouped LLM-generated topic labels into seven categories: “General Knowledge”, “Health”, “Internet, Technology, Media”, “Shopping”, “Lifestyle”, “Business, Finance, Employment” and “Covid”. “Covid” is a special category we defined, which includes all queries that mention “Covid” or “Coronavirus.”

\begin{table}[!h]
\centering
\begin{tabular}{lr}
\hline
Style & count \\ 
\hline
Question & 7560 \\ 
 Statement & 2360 \\ 
Navigational & 1452 \\ 
   \hline
\end{tabular}
\caption{Style of Search Queries}
\label{si-tab:query-style}
\end{table}

\begin{table}[!hb]
\centering
\begin{tabular}{lr}
 \hline
Topic & Count \\ 
  \hline
Internet, Technology, Media & 3421 \\ 
General Knowledge & 2467 \\ 
Business, Finance,Employment & 2137\\ 
 Lifestyle & 2017 \\ 
Health & 808 \\ 
 Shopping & 448 \\ 
 Covid & 74 \\ 
   \hline
\end{tabular}
\caption{Topic of Search Queries}
\label{si-tab:query-topic}
\end{table}

\newpage

\subsection{Predicting the Appearance of GenAI Search Responses}
We evaluated the predictive power of three feature sets---country, query style, and query topic---in predicting whether a Google search result would include a GenAI (AI Overview) response. We considered it as a binary classification task with “0” representing “No AI Overview” and “1” representing “Has AI Overview.” To achieve this, we trained three Random Forest (RF) models, each using one feature set, as well as a full RF model using all three feature sets. We chose Random Forest because it is well-suited to supervised learning tasks, handles categorical data well, and provides reliable estimates of feature importance. Model performance was evaluated using the Area Under the Receiver Operating Characteristic Curve (AUC), a metric that captures the trade-off between true positive and false positive rates and is threshold-independent. To ensure robustness in performance estimation, we trained and evaluated each model using 5-fold cross-validation (100 repeats), which generated 500 AUC scores per model. Such analysis allowed us to compare the performance of each feature set when using the same type of model and to gain insight into which type of features are more important when predicting the appearance of a GenAI response. 

To further investigate feature importance, we fitted the model on all features (i.e., the full all-feature model) and calculated the Gini importance score for each feature. Gini importance provides a measure of how much each feature reduces node impurity in the RF model; a feature with a higher Gini importance score is more influential in the classification task.

Figure 1C in the main text displays the Gini importance scores for all features, and Figure 1D in the main text presents the performance comparison of the RF models based on their AUC scores.

\newpage

\section{Experiment}
 The experiment was conducted with 4,927 participants from March 20 to April 2, 2024. We preregistered our hypotheses, experiment design, sample size, and analysis plan. Participants provided informed consent, were told they might be given fictitious information at the beginning of the experiment, and were debriefed after the experiment. The study was reviewed and approved by the MIT Institutional Review Board (IRB) (E-5570). The data, preregistration, and survey are available online at \href{https://osf.io/2du7x/}{https://osf.io/2du7x/}. 

\subsection{Experiment Design}\label{experimentdesign}
The experiment employs a mixed design. Participants were randomized into one of the following experiment groups: GenAI, traditional search, explanation, reference, uncertainty highlighting, and social feedback. Participants who were assigned to the latter three groups were again randomized into a baseline version or a variation version of their assigned treatment, at search task level. 

\begin{itemize}
    \item \textbf{GenAI, n = 814.} The GenAI group were given generative responses, with no trust design cues, and were informed that the responses were created by generative AI. 
    \item \textbf{Traditional search, n = 842.} The traditional search group were given generative responses presented in the Google featured snippet UI, with no trust design cues, and were informed that the responses were not created by generative AI.
    \item \textbf{Explanation treatment, n = 812.} The explanation treatment group were first provided with a short explanation paragraph (which is an introduction to, and the pros and cons of, Google AI Overviews). Then participants were given generative responses and informed that the responses were created by generative AI.
    \item \textbf{Reference treatment, n = 785.} The reference treatment group were given generative responses and clickable reference links, and were informed that the responses were created by generative AI. They were then randomized into a baseline version or a variation version at search task level. In the baseline version, all reference links in the responses were valid. In the variation version, some references in the responses were broken, invalid or hallucinated.
    \item \textbf{Uncertainty highlighting treatment, n = 817.} The uncertainty highlighting treatment group were given generative responses with color-coded uncertainty highlighting that indicates statements of high/low uncertainty, and were informed that the responses were created by generative AI. They were then randomized into a baseline version or a variation version at search task level. In the baseline version, the responses highlighted both high and low uncertainty statements. In the variation version, the responses only highlighted high uncertainty statements.
    \item \textbf{Social feedback treatment, n = 857.} The social feedback treatment group were given generative responses and social feedback regarding the helpfulness of those responses. They were informed that the responses were created by generative AI. They were then randomized into a baseline version or a variation version at search task level. In the baseline version, the feedback was positive, displaying a fraction of “users who found this helpful” chosen randomly from a uniform distribution between 65\% and 95\%. In the variation version, the feedback was negative, displaying a fraction of “users who found this helpful” chosen randomly from a uniform distribution between 5\% and 35\%.
\end{itemize}

\subsection{Participants}
We recruited a 4,927 person sample (pre-registered n=5,000) representative of the adult population of the U.S. on \href{https://www.prolific.com}{Prolific}, an online crowdsourcing platform that has been shown to have participants that are more diverse, less dishonest, and produce good quality data\cite{Peer017prolific}. Our sample matches the US distribution of age, gender, and ethnicity.

\subsection{Procedure}
Recruited Prolific participants were directed to complete the experiment on \href{https://www.qualtrics.com}{Qualtrics}. They first answered demographic questions, and then were given a brief introduction to Google AI Overviews and a description of their assigned treatment. They next reviewed 9 pre-defined search tasks and generative search responses to the search tasks collected from Google AI Overviews (the ``stimuli''), each on a separate page, in random order. For each search task and response, they rated how much they agreed with the following six statements: (1)-(5) ``How do you consider the response to this question? I think this response is accurate / believable / biased / complete / trustworthy'' and (6) ``How much do you agree with the following statement? I am willing to share this response with a friend who is interested in this question/topic.'' We use their ratings to construct two outcome variables. After finishing all the search tasks, they were given stimuli modification details and received a fixed payment of \$2.40. On average, participants spent 13 minutes and 40 seconds on the experiment (SD = 8.33 minutes; Median = 11.15 minutes). 

\newpage 

\subsection{Stimuli Collection}\label{stimulicollect}
Stimuli collection was conducted in January, 2024. To collect generative responses, we used Google AI Overviews (It was called Google Search Generative Experience (SGE) at the time of stimuli collection and experiment). This generative search engine has the largest potential user base given its integration with Google Search, and is powered by Google's state-of-the-art large language models (LLMs).

We pre-defined the following 9 search queries to create experiment stimuli:
\begin{enumerate}
\itemsep0em 
    \item Will increased gun control lead to more or less crime?
    \item Is implementing universal healthcare in the US negative or positive for the health of citizens?
    \item Does more immigration take jobs from or create jobs for US citizens?
    \item Does increased civilian oversight of police lead to more or less crime?
    \item Will raising taxes on wealthy individuals within the US decrease or increase tax revenue?
    \item Government policy to fight inflation (\textit{inflation})
    \item Jobs replaced by ai (\textit{AI and employment})
    \item How often do adults need vaccines (\textit{vaccine})
    \item Harmful effects of climate change (\textit{climate change})
\end{enumerate}

Query (1)-(5) are debate questions on policy issues, modified from topics used in \cite{rand2023numericalab} (pretesting conducted in \cite{rand2023numericalab} indicates that these topics are highly polarizing for both Democrats and Republicans). Query (6)-(9) are regular questions. To collect these questions, we first used keywords from \href{https://www.pewresearch.org/politics/2023/06/21/inflation-health-costs-partisan-cooperation-among-the-nations-top-problems/}{a list of US top problems surveyed by Pew Research Center}, and collected relevant queries from Google's ``People also ask''. We searched these 9 queries in Google and took screenshots of their corresponding generative responses provided by Google AI Overviews. To ensure our responses were not personalized, we used a new Google account with no search history, disabled the personalization setting, and searched in a private window. We then edited the presentation of responses to create the experimental stimuli as described in \ref{si:stimuli-edit}. 

\newpage
\begin{figure}[!ht]
    \centering
    \begin{subfigure}[b]{0.45\textwidth}
        \centering
        \includegraphics[width=\textwidth]{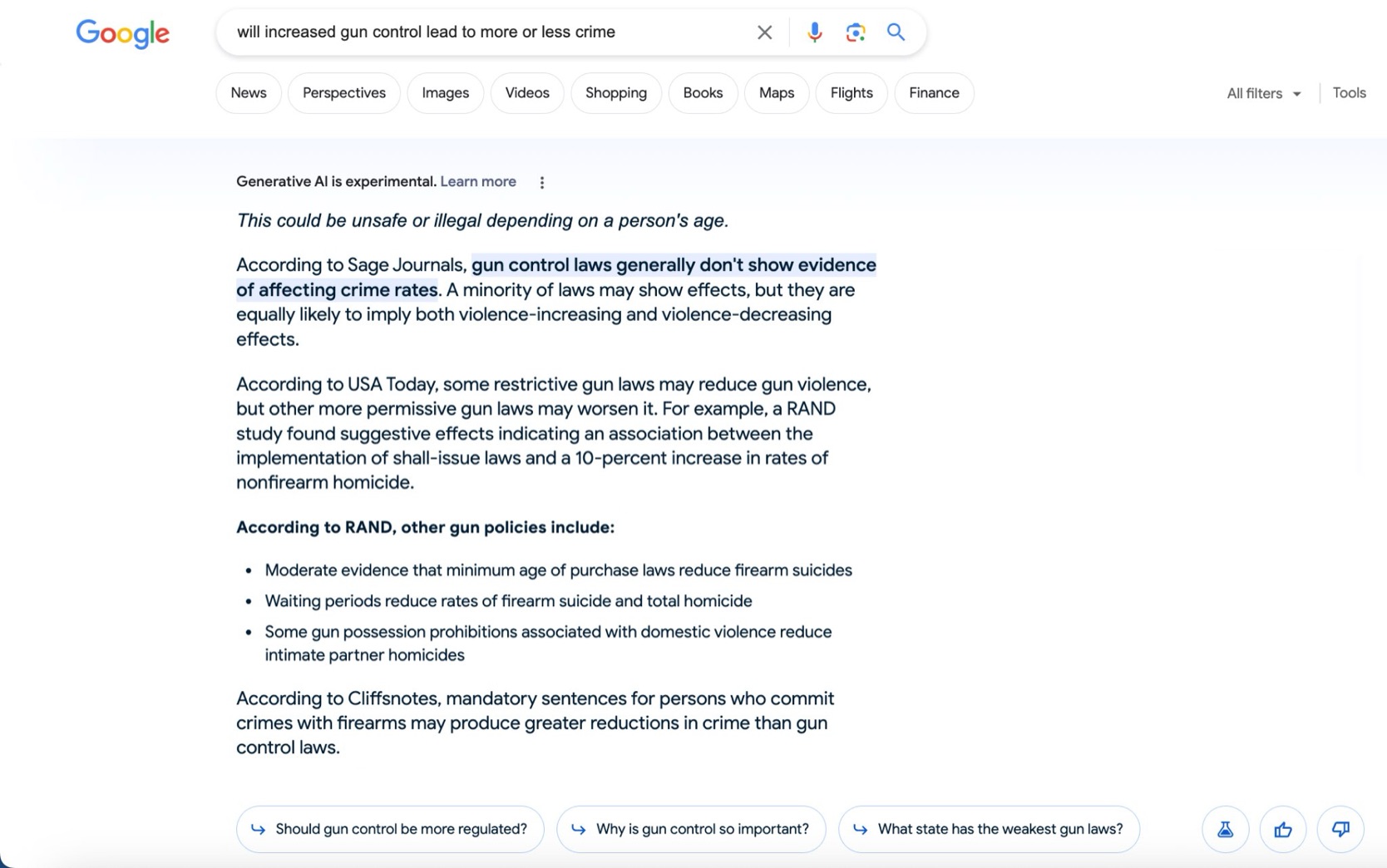}
        \caption{GenAI}
    \end{subfigure}
    \hfill
    \begin{subfigure}[b]{0.45\textwidth}
        \centering
        \includegraphics[width=\textwidth]{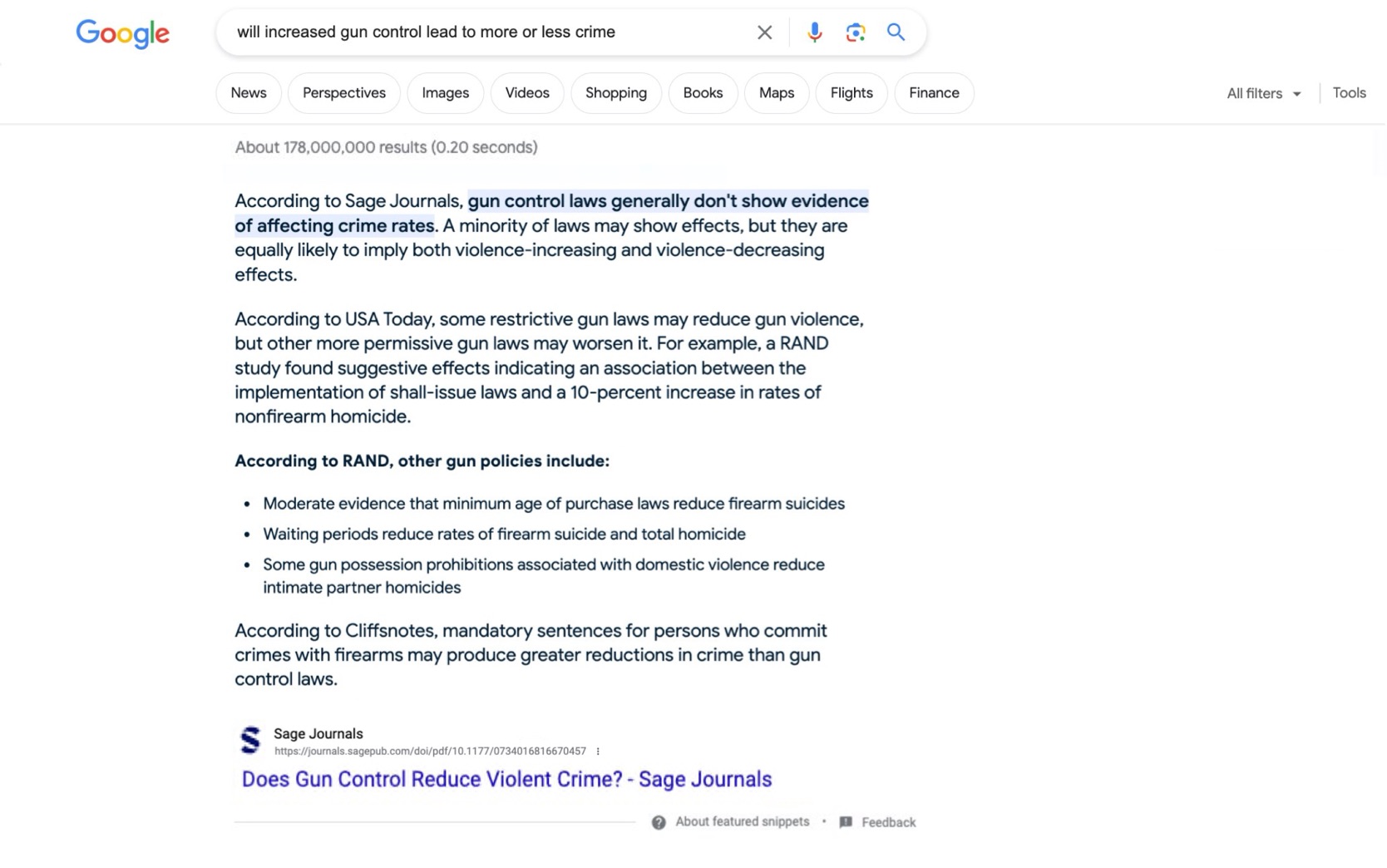}
        \caption{Traditional Search}
    \end{subfigure}
    \vskip\baselineskip
    \begin{subfigure}[b]{0.45\textwidth}
        \centering
        \includegraphics[width=\textwidth]{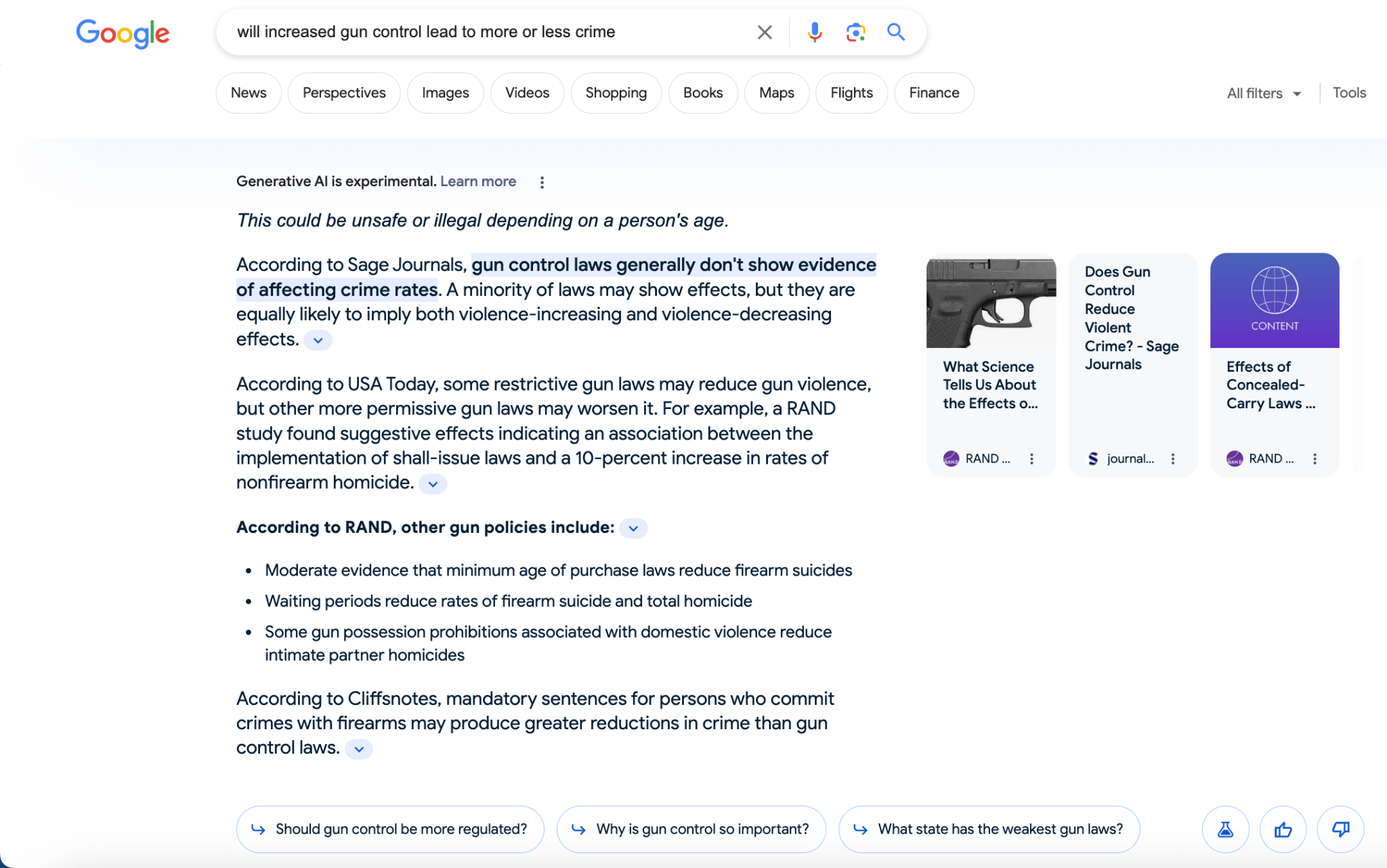}
        \caption{Reference}
    \end{subfigure}
    \hfill
    \begin{subfigure}[b]{0.45\textwidth}
        \centering
        \includegraphics[width=\textwidth]{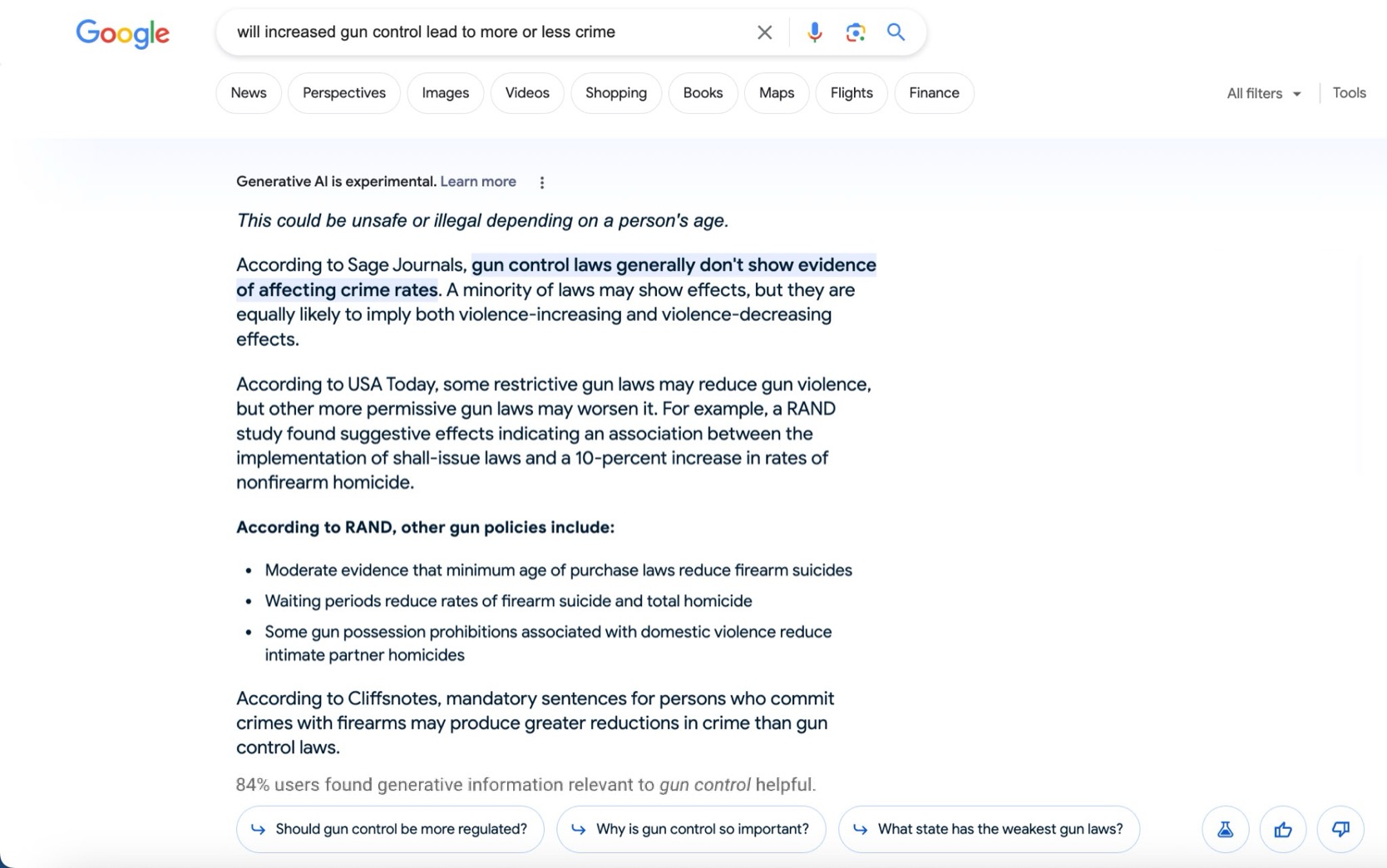}
        \caption{Positive Feedback}
    \end{subfigure}
    \vskip\baselineskip
    \begin{subfigure}[b]{0.45\textwidth}
        \centering
        \includegraphics[width=\textwidth]{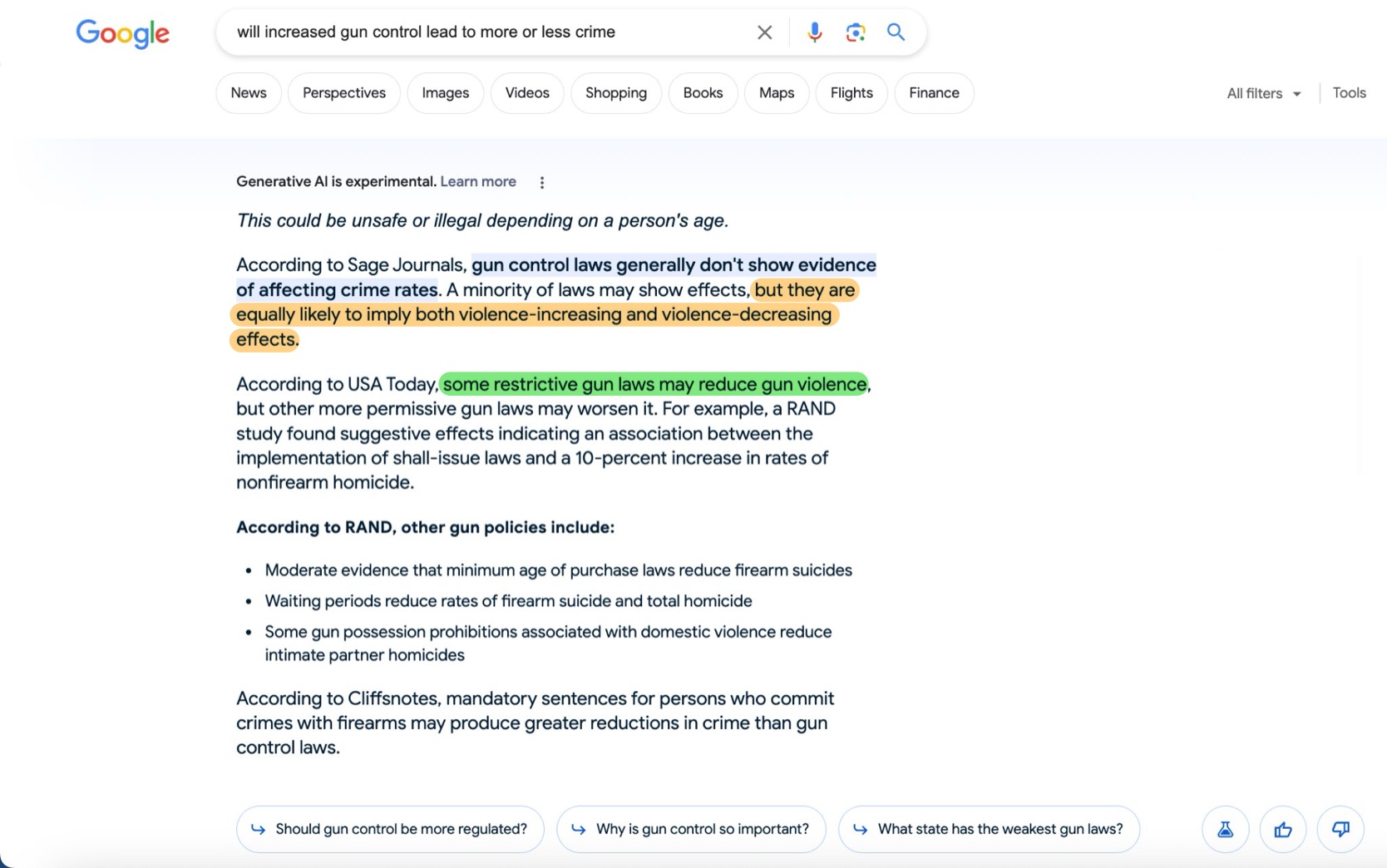}
        \caption{High and Low Uncertainty Highlighting}
    \end{subfigure}
    \hfill
    \begin{subfigure}[b]{0.45\textwidth}
        \centering
        \includegraphics[width=\textwidth]{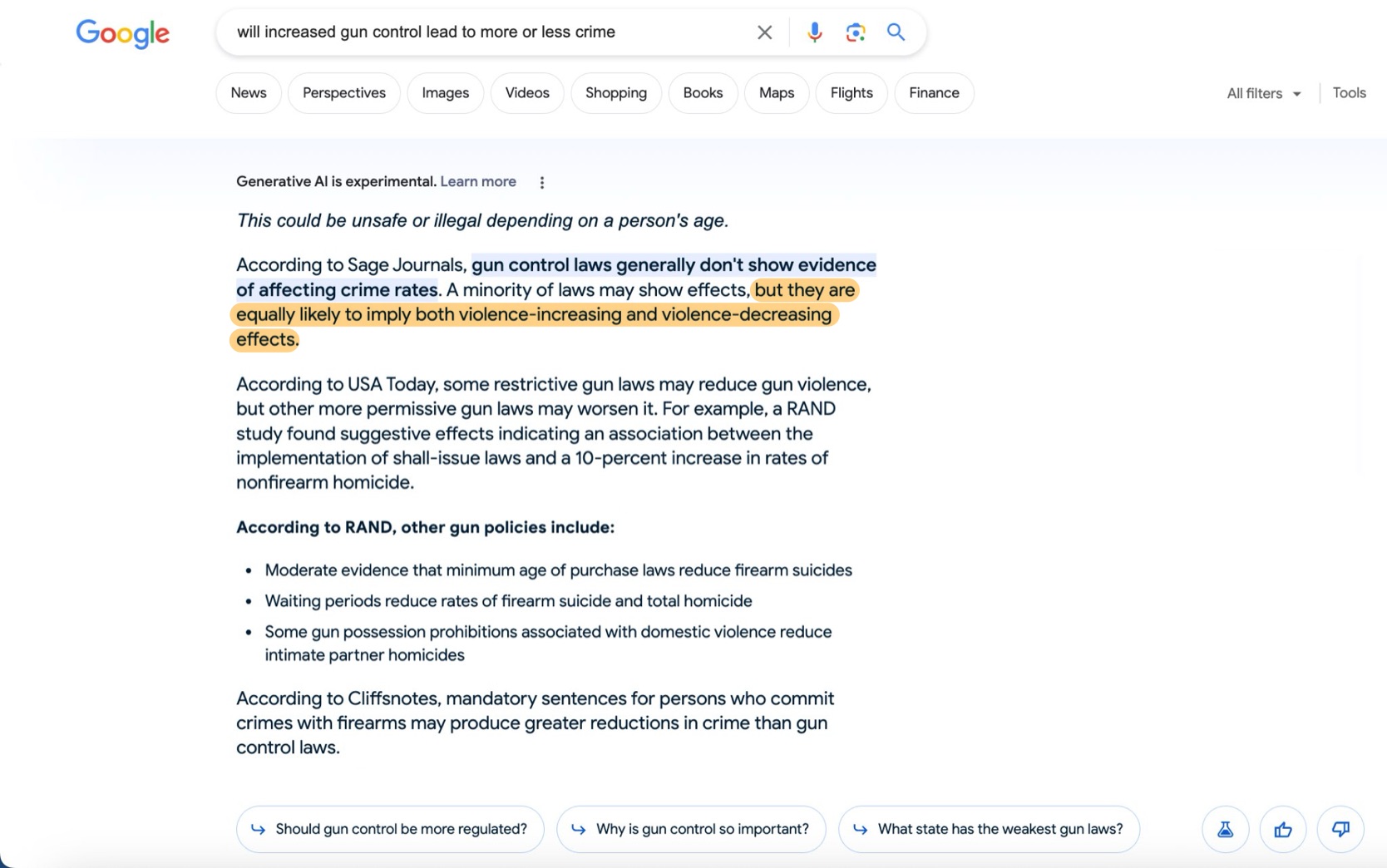}
        \caption{High Uncertainty Highlighting Only}
    \end{subfigure}
    \caption{Stimuli Examples}
    \label{si-fig:stimuli-example}
\end{figure}

\newpage

\subsection{Stimuli Editing}\label{si:stimuli-edit}
The original responses from Google AI Overviews consisted of two or three short paragraphs with references. We manipulated the presentation by manually adding or removing trust design cues or presenting the response in the traditional Google featured snippet UI format to create the stimuli. We did not make any changes to the response text itself. Figure \ref{si-fig:stimuli-example} displays stimuli examples experienced by participants. 

\begin{figure}[!hb]
    \centering
    \includegraphics[width=0.8\linewidth]{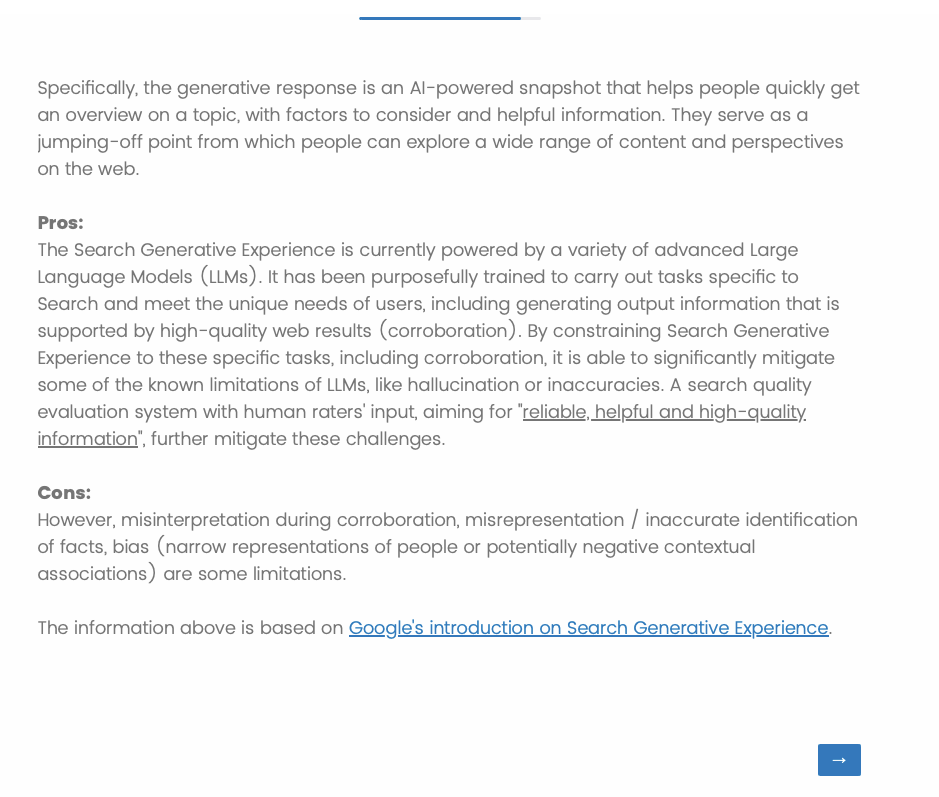}
    \caption{A Brief Introduction Shown to The Explanation Group}
    \label{si-fig:explain_text}
\end{figure}

\begin{itemize}
    \item GenAI stimuli are the collected generative responses with all references removed. 
    
    \item To create the traditional search group stimuli, we displayed the text of the generative responses collected from Google AI Overviews in the traditional Google featured snippet UI format. 
    
    \item Reference group stimuli were simply the original generative responses. Clickable references included in the original responses were provided with the stimuli. In the baseline version, no changes were made to the original generative responses. In the variation version, we replaced one reference with a broken, invalid or hallucinated one in 7 of the 9 tasks. We excluded changing responses in tasks 6 and 8 because the original references provided in response to these two search tasks were already hallucinated. More details are discussed in \ref{si:q68}.

    \item The uncertainty highlighting group stimuli were the generative responses with all references removed. We added orange and/or green highlighting: we added green highlighting to statements with valid references; we picked claims that were verified to have conflicting evidence online to add orange highlighting. Participants were told that green highlighting means ``Google Search found content that's likely similar to the statement.'' and orange highlighting means ``Google Search found content that's likely different from the statement, or it didn't find relevant content.'' In the baseline version, the response included both orange and green highlighting. In the variation version, the response included only orange highlighting, i.e., high uncertainty highlighting only.
    
    \item The social feedback group stimuli were the generative responses with all references removed. In addition, we added one sentence at the end of every response: ``X\% of users found generative information relevant to \textit{the topic} helpful,'' with X being a random integer. In the baseline version, X was an integer randomly drawn from a uniform distribution between 65 and 95. In the variation version, X was an integer randomly drawn from uniform distribution between 5 and 35.

    \item The explanation group stimuli were generative responses with all references removed. In addition, the explanation group were shown a brief introduction (see Figure \ref{si-fig:explain_text}), modified from Google's explanation on what Search Generative Experience is and its pros and cons, and a link to \href{https://static.googleusercontent.com/media/www.google.com/en//search/howsearchworks/google-about-SGE.pdf}{Google’s official introduction document on the Search Generative Experience}.
\end{itemize}
\newpage

\newpage

\section{Data and Descriptive Statistics}
We pre-registered a target sample size of 5,000, and we finally recruited and approved payment to 4,993 participants. We did not receive a complete response from 47 of the participants, leaving us with a 4,946 person sample. Following our pre-registration, we further excluded 19 participants who spent less than 3 minutes on the survey. 

The final sample is a US representative sample of 4,927 Prolific participants that matches the US population in terms of age, gender, and ethnicity. Participants are required to be at least 18 years old and US-based to be included in the experiment. The sample is 50.6\% female, and has an average age of 42.2 years (Female proportion and average age are calculated using the demographic information of recruited participants provided by Prolific. In our experiment we only offered age range options for participants to select when asking about their age, and offered ``Non-binary/third gender'' and ``Prefer not to answer'' options when asking about their gender. We only used participants' survey responses, not the demographic information provided by Prolific, in our analysis.) We ended up having more participants falling within the 25-44 age bracket and less above 65 because the platform encountered difficulties in filling less common demographics with the large sample size we targeted. Table \ref{si-tab:descriptive} and \ref{si-tab:usecase} display descriptive statistics of our sample and a comparison with the US population.

\subsection{Pre-treatment Covariates}
We pre-registered eight categorical pre-treatment covariates: age, gender, ethnicity, education degree, career, political leaning, frequency of generative AI use, and frequency of generative search engine use. We controlled for these covariates in our analysis\cite{gerber2012fieldexperiment}. 

We also asked participants what they use generative AI and generative search engines for. Among the 84.98\% of the participants who said they had used generative AI before, information search was the most popular use case (62.91\%). In addition, 67.83\% of the sample had used generative search engines before, and 82.23\% of them had used generative search engines to find answers to questions (see Table \ref{si-tab:usecase}). 

\begin{table}[htbp!]
    \centering
    \small
    \caption{Descriptive Statistics (\%).}
    \begin{minipage}{\textwidth}
    \makebox[\linewidth][c]{%
        \begin{tabular}{lllcc}
        \hline
    Variable& Value(analysis)& Value(survey)& Sample & US adult population\footnote{Based on 2022 US census data from \href{https://data.census.gov}{https://data.census.gov}}\\
            \hline
            Age & & 18 - 24 & 12.85 & 12.98\\
            & & 25 - 34 & 24.54 & 17.47\\
            & & 35 - 44 & 21.72 & 16.92 \\
           & & 45 - 54 & 17.43 & 15.72\\
           & & 55 - 64 & 14.55 & 16.52 \\
           & & 65 or older & 8.91 & 22.01\\
            \hline
            Gender & & Female & 49.00 & 51.21\\
           & & Male & 48.65 & 48.79\\
          &  & Non-binary/third gender & 2.11 & /\\
          &  & Prefer not to answer & 0.24 & /\\
            \hline
            Ethnicity\footnote{Alone or in combination with one or more other races}& & White & 73.98 & 71.0\\
           & & Black & 13.44 & 14.2 \\
          &  & Asian & 7.85 & 7.2\\
          &  & Hispanic/Latino & 12.36 & 18.7 \\
          &  & Native Americans & 2.39 & 2.9 \\
           & & Pacific Islander & 0.43 & 0.5 \\
          &  & Others  & 1.20 & / \\
            \hline
            Career & Work in tech & Technology & 16.07&\\
          & Not in tech  & Public services & 6.35& \\
          &  & Business\&Finance & 10.70&\\
           & & Education & 8.65&\\
            && Healthcare & 9.30&\\
            && Arts\&Media & 9.34&\\
            & &Manufacturing & 7.25&\\
            & &Others & 39.15&\\
            \hline
            Education & No college degree & Less than high school & 0.67&\\
           & & High school graduate & 12.18&\\
           & & Some college & 20.76&\\
          & With college degree  & 2 year degree & 9.50&\\
          &  & 4 year degree & 39.82&\\
          &  & Professional degree & 15.02&\\
           & & Doctorate & 2.05&\\
            \hline
            Political leaning & Rep & Strongly republican & 4.51&\\
          &  & Republican & 9.01&\\
          &  & Lean republican & 10.15&\\
          &Neutral  & Neutral & 18.61&\\
          & Dem  & Lean democratic & 16.44&\\
          &  & Democratic & 21.55&\\
          &  & Strongly democratic & 19.73&\\
            \hline
            Generative AI usage &Never & Never & 15.02&\\
           &Low & Less than once a week & 32.64&\\
           & & Once a week & 13.80&\\
          &  & 2-3 days a week & 18.77&\\
           &High & 4-6 days a week & 10.01&\\
           & & Daily & 9.76&\\
            \hline
            Generative search usage & Never & Never & 32.17&\\
           &Low & Less than once a week & 26.47&\\
           & & Once a week & 8.95&\\
           & & 2-3 days a week & 13.68&\\
           &High & 4-6 days a week & 8.02&\\
           & & Daily & 10.72&\\
    \hline
\end{tabular}%
}
\label{si-tab:descriptive}
\end{minipage}
\end{table}

\begin{table}[htbp!]
    \centering
    \caption{Use Cases of GenAI and Generative Search Engines.}
    \begin{subtable}[t]{\linewidth}
        \centering
        \begin{tabular}{lr}
        \hline
         & Proportion(\%) \footnote{Proportion of the sample of participants who indicated they haved used genAI before. }\\
        \hline
        Information Search & 62.91\\
        Content generation/writing assistance & 51.40\\
        Personal assistance/recommendation & 51.25\\
        Learning & 48.15\\
        Work & 40.12\\
        Artwork/music creation & 19.25\\
        Other & 6.35\\
        \hline
        \end{tabular}
        \caption{Use Cases of GenAI}
        \label{si-tab:genai-usecase}
    \end{subtable}
    \vspace{0.5cm}
    \begin{subtable}[t]{\linewidth}
        \centering
        \begin{tabular}{lr}
            \hline
             & Proportion(\%) \footnote{Proportion of the sample of participants who indicated they haved used generative search engines before. }\\
            \hline
            Find answer to a question & 82.23\\
            Find websites/information on a topic but I don’t have a specific site in mind & 46.44\\
            Get recommendations & 46.05\\
            Find products, compare prices, make purchases & 32.23\\
            Reach a particular website that I already have in mind & 17.77\\
            Other & 4.25\\
            \hline
            \end{tabular}
            \caption{Use Cases of Generative Search Engines}
        \label{si-tab:aisearch-usecase}
    \end{subtable}
    \label{si-tab:usecase}
\end{table}

\newpage

\subsection{Outcome Variables}
Our primary outcome variables are the 5-item trust rating and the willingness to share rating. We constructed the 5-item trust rating as a combination of accuracy, believability, (un)biasedness, completeness, and trustworthiness. Specifically, after experiencing every search task, participants were asked ``How do you consider the response to this question?'' and rated the following statements on a 7-point likert scale, with 1 corresponding to ``strongly disagree'' and 7 corresponding to ``strongly agree''.  We averaged the 5 ratings at the participant-search task level to calculate the 5-item trust rating. 
\begin{enumerate}
    \item I think this response is \textit{accurate}.
    \item I think this response is \textit{believable}.
    \item I think this response is \textit{biased}.
    \item I think this response is \textit{complete}.
    \item I think this response is \textit{trustworthy}.
\end{enumerate}

We reversed the Likert scale rating for biasedness to measure trust, rather than lack of trust.

We also asked participants to rate ``How much do you agree with the following statement? I am willing to share this response with a friend who is interested in this question/topic,'' on the same 7-point likert scale, and used the response as the willingness to share rating. 

\newpage

\subsection{Reliability and Validity of 5-item Trust and Willingness to Share}
Previous research examined people's trust in, or perceived credibility of, web-based information at various levels, including message, source, sponsor, and channel/media, and commonly operationalized it as a multi-dimensional concept which comprises separate items such as completeness, trustworthiness, authenticity, etc\cite{appelman2016measuring, flanagin2007role, kiousis2001channel, metzger2003credibility}.  We follow the message credibility literature to construct a trust measure, as our focus is on people's perception of generative text rather than source or channel, which are less salient and prominent factors in the generative search context. Specifically, we selected five items — believability, accuracy, trustworthiness, bias, and completeness — to measure trust, following\cite{flanagin2000perceptions,flanagin2007role} in which the authors summarized previous studies and proposed these items as the most important for assessing message credibility. We consider these five items appropriate because they logically contain and reflect the concept of trust. Additionally, including too many items would be less feasible due the risk of participant fatigue, while too few items might have failed to capture perceived trust comprehensively.  

To elicit participants' perceptions of generative responses based on these trust items, we asked them to rate the statements: ``I think this response is accurate/believable/biased/complete/trustworthy.'' These statements explicitly reference ``response'' to avoid ambiguity and ensure clarity that we are asking their opinion of the search response. Participants rated these statements on a 7-point Likert scale, with 1 corresponding to ``strongly disagree'' and 7 corresponding to ``strongly agree.'' The Likert scale was devised and is widely used to provide a alidated measure of people's attitudes, and 7-point Likert scales are considered to offer a broader spectrum of choices than the 5-point version, allowing participants greater independence to select a more precise level of agreement, rather than a more distant option \cite{joshi2015likert}.

After data collection, we followed our pre-registration to exclude responses from participants who did not complete the experiment or who completed it in less than 3 minutes. We assessed the reliability and validity of the 5-item trust rating, and validated that the five items are good indicators that reliably represent the latent trust factor. Specifically, to measure reliability and internal consistency, i.e. how closely the five observed trust items are as a group and how well they measure the same latent variable, we used Cronbach's alpha, following the methodology of \cite{appelman2016measuring, flanagin2007role}. Cronbach's alpha evaluates the amount of shared variance among the items relative to the overall variance \cite{cronbach1951coefficient}. Table \ref{tab:cronbach-a} presents the Cronbach's alpha values. For each search task, the cronbach value is based on all partcipants' five ratings of that search task response. All alpha values are greater than 0.87, passing the 0.8 rule-of-thumb, suggesting high reliability \cite{flanagin2007role}. 

To verify construct validity, we conducted a confirmatory factor analysis (CFA). The individual factor loadings indicate the direction and strength of the relationship between observable variables and the latent construct. Table \ref{tab:cfa} presents the results of the CFA on our five observed trust items. For all search tasks, all the factor loadings were positive, significant, and exceeded the 0.4 cutoff as recommended in \cite{gefen2000structural}, demonstrating a strong relationship between the five items and the underlying trust factor. Furthermore, high t-statistics suggest that the five observed items converge to measure the same underlying trust construct and suggest highly significant contributions to the construct, supporting their convergent validity.

In addition, we asked participants to report their level of agreement with the statement, ``I am willing to share this response with a friend who is interested in this question/topic,'' using the same 7-point Likert scale. Such willingness to share ratings measure an actionable intention and elicit perceived trust in the response. The willingness to share has been widely used in previous work to measure people's perceptions of and belief in of online (mis)information \cite{pennycook2021practical}. Evidence suggests a meaningful relationship between the self-reported willingness to share of crowd-sourcing platform users and their actual levels of sharing of online content \cite{mosleh2020self}. In our context, we consider sharing an AI-generated response with a friend as a proactive behavior and an indicator of a higher level of trust and confidence in the response. That is, people are more willing to share the response if they consider it credible and valuable enough to recommend to a friend they know, and they are unlikely to share content they deem untrustworthy. While the willingness to share likely measures both trust and participants’ perceptions of the value of search information for others, people are more willing to share responses they find credible. Sharing intentions also indicate the extent to which GenAI may diffuse from person to person in human information seeking and problem-solving tasks and therefore the likelihood that generative content will spread through human social networks.

\begin{table}[hbtp!]
    \centering
    \caption{Cronbach's Alpha of 5-item Trust Rating. }
    \begin{tabular}{ccccc}
    \hline
    Gun control & Healthcare & Immigration & Civilian oversight&Tax\\
    0.9 & 0.873 & 0.905 & 0.89 & 0.871 \\
    \hline
    Inflation&Job replaced by AI&Vaccine&Climate change&\\
     0.875 & 0.88 & 0.888 & 0.883&\\
    \hline
\end{tabular}
\label{tab:cronbach-a}
\end{table}

\begin{table}[hbtp!]
\renewcommand{\arraystretch}{0.8}
\caption{Confirmatory Factor Analysis on the Five Observed Trust Items. For every search task, standardized factor loadings are all positive, significant and above the 0.4 cutoff. The individual factor loadings and the high t-statistics suggest good construct validity and convergent validity.}
    \centering
        \begin{tabular}{llcc}
        \hline
        Task & Item & Factor loading & t statistics\\
        \hline
        Gun control & believable & $0.913^{***}$ & 305.399\\
         & complete & $0.755^{***}$ & 115.688\\
         & trustworthy & $0.907^{***}$ & 291.659\\
         & accurate & $0.941^{***}$ & 386.627\\
         & unbiased & $0.533^{***}$ & 50.291\\
         \addlinespace[1.5ex]
        Healthcare & believable & $0.890^{***}$ & 235.361\\
         & complete & $0.716^{***}$ & 95.455\\
         & trustworthy & $0.872^{***}$ & 210.444\\
         & accurate & $0.925^{***}$ & 296.649\\
         & unbiased & $0.511^{***}$ & 46.308\\
        \addlinespace[1.5ex]
        Immigration & believable & $0.919^{***}$ & 332.516\\
         & complete & $0.748^{***}$ & 112.334\\
         & trustworthy & $0.911^{***}$ & 308.436\\
         & accurate & $0.948^{***}$ & 432.532\\
         & unbiased & $0.563^{***}$ & 55.896\\
        \addlinespace[1.5ex]
        Civilian oversight & believable & $0.873^{***}$ & 213.847\\
         & complete & $0.775^{***}$ & 123.799\\
         & trustworthy & $0.895^{***}$ & 247.775\\
         & accurate & $0.921^{***}$ & 296.312\\
         & unbiased & $0.512^{***}$ & 46.495\\
        \addlinespace[1.5ex]
        Tax & believable & $0.878^{***}$ & 214.548\\
         & complete & $0.728^{***}$ & 99.898\\
         & trustworthy & $0.881^{***}$ & 217.614\\
         & accurate & $0.913^{***}$ & 266.068\\
         & unbiased & $0.483^{***}$ & 42.086\\
        \addlinespace[1.5ex]
        Inflation & believable & $0.912^{***}$ & 293.646\\
         & complete & $0.732^{***}$ & 103.501\\
         & trustworthy & $0.897^{***}$ & 262.309\\
         & accurate & $0.932^{***}$ & 342.351\\
         & unbiased & $0.415^{***}$ & 33.936\\
        \addlinespace[1.5ex]
        AI and employment & believable & $0.882^{***}$ & 228.372\\
         & complete & $0.741^{***}$ & 106.622\\
         & trustworthy & $0.893^{***}$ & 245.383\\
         & accurate & $0.928^{***}$ & 312.313\\
         & unbiased & $0.470^{***}$ & 40.662\\
        \addlinespace[1.5ex]
        Vaccine & believable & $0.903^{***}$ & 280.180\\
         & complete & $0.759^{***}$ & 117.090\\
         & trustworthy & $0.908^{***}$ & 292.229\\
         & accurate & $0.940^{***}$ & 377.750\\
         & unbiased & $0.511^{***}$ & 46.798\\
        \addlinespace[1.5ex]
        Climate change & believable & $0.920^{***}$ & 323.793\\
         & complete & $0.681^{***}$ & 84.745\\
         & trustworthy & $0.895^{***}$ & 265.040\\
         & accurate & $0.946^{***}$ & 400.211\\
         & unbiased & $0.560^{***}$ & 55.080\\
        \hline
        \multicolumn{3}{l}{\rule{0pt}{1em}* p $<$ 0.05, ** p $<$ 0.01, *** p $<$ 0.001}\\
        \end{tabular}\label{tab:cfa}
\end{table}

\newpage

\section{GenAI Treatment Selection Motivation}
\textbf{Reference.} Current studies have proposed using references (external knowledge) to provide sources and corrections in the generation process, for building more verifiable language models \cite{huang2023citation, zhang2023sirens}. Popular generative search engines such as Google AI Overviews and Microsoft’s AI-powered Bing Chat also offer in-line references that users can use to fact check and dig deeper. However, research has pointed out that they are not properly sourced frequently \cite{li2024generativeaisearchengines, liu2023evaluategse}. 

\textbf{Uncertainty Highlighting.} Another aspect of AI systems that researchers strive to enhance is their transparency regarding uncertainty or potential errors \cite{xiong2024llms,zhang2023sirens}. Google’s Bard/Gemini introduced a color-coded highlighting feature to help users identify statements that have supportive or contradictory evidence online. Previous studies verified the efficiency of such features in directing workers attention to content needing inspection \cite{gosline2024friction}, assisting programmers to spot problematic LLM-generated code \cite{vasconcelos2023coding}, and alerting consumers to potentially misleading information when searching for products with generative search engines \cite{spatharioti2023microsoft}. 

\textbf{Social Feedback.} Social feedback is a common signal that people use to calibrate their trust and has been studied extensively in many fields, including people's trust in online information. However, the conventional social feedback is unrealistic in the generative AI context because every response is unique. Therefore, we introduced the social feedback in the form of generative search helpfulness on a topic, considering that responses to the same topic/query are much similar. 

\textbf{Explanation.} By providing explanations of AI systems that are understandable to humans is also important for developing people's trust in AI. We examine its effectiveness in the generative search context. In our experiment we provided participants with additional information on the general purpose, benefits, and limitations of generative search (it can be viewed as a high-level explainability). 

\newpage

\section{Treatment Randomization}
We implemented simple randomization on Qualtrics to assign participants into one of the five treatment groups. We used the Fisher-Yates shuffle algorithm to implement search task order randomization and within-subjects treatment variation assignment, to mitigate order effects. A chi-squared test of independence between search task and order returned  $\chi^2$ = 69.877, df = 64, p-value = 0.287, demonstrating independence of the search task and order. An independence test between treatment variation and order returned $\chi^2$ = 6.930, df = 8, p-value = 0.544, demonstrating independence of the treatment variation and the order. Moreover, the two randomization processes were implemented independently, and separately for each participant. A chi-squared test of independence between search task and treatment variation returned $\chi^2$= 5.507, df = 8, p-value = 0.702. 

\subsection{Treatment Group Size}\label{si:group-size}
Participants were randomly assigned to one of the treatment groups with equal probability. Table \ref{si-tab:group-size} shows the number of participants in each group. The group sizes are generally balanced but not identical because we used simple randomization. We found no evidence of imbalance on pre-treatment covariates or differential attrition (see \ref{si:attrition} and \ref{si:cov-balance} for more details).

\begin{table}[htbp!]
    \begin{minipage}{\textwidth}
        \caption{Group Sizes}
                \centering
        \begin{tabular}{lrr}
        \toprule
            & N & \%\\
        \midrule
        GenAI & 814 & 16.52\\
        Traditional search & 842 & 17.09\\
        Reference\footnote{The reference treatment group has a smaller sample size, and is possibly due to its slightly higher attrition rate. We discuss in \ref{si:attrition} that this attrition is random.}  & 785 & 15.93\\
        Uncertainty highlighting & 817 & 16.58\\
        Social feedback & 857 & 17.39\\
        Explanation & 812 & 16.48\\
        \midrule
        Total & 4927 & 100.00\\
        \bottomrule
    \end{tabular}
    \label{si-tab:group-size}
  \end{minipage}
\end{table}

\newpage

\subsection{Attrition}\label{si:attrition}
The overall dropout rate after being assigned to experiment groups is 2.62\%, and the reference treatment group has a slightly higher dropout rate of 4.24\%. This is likely due to randomness and does not affect our results: we tested for differential attrition by running a logistic regression to predict dropout rates using treatment assignment, and none of the treatment groups significantly predicted dropout at the 0.05 significance level (see Table \ref{si-tab:attrition}). This means participants who were assigned to any treatment group were no more likely to drop out than participants in the GenAI group. Moreover, reference treatment group attrition can not be predicted from age, gender, ethnicity, career, education level, political leaning, frequency of generative AI use, or frequency of generative search engines use (F-statistic: 0.778 on 45 and 780 DF,  p-value: 0.853). In addition, we show our final sample is balanced in terms of pre-treatment covariates in \ref{si:cov-balance} (a selective attrition test). 

\begin{table}
\centering
\caption{Attrition Check}
\begin{tabular}[hb!]{lcc}
\toprule
\multicolumn{1}{c}{ } & \multicolumn{2}{c}{DV: dropout} \\
\cmidrule(l{3pt}r{3pt}){2-3}
  & Coefficient & p-value\\
\midrule
Intercept (GenAI) & \num{-3.665} & \num{<0.001}\\
Explanation & \num{-0.203} & \num{0.539}\\
Feedback & \num{-0.202} & \num{0.535}\\
Highlighting & \num{0.215} & \num{0.470}\\
Reference & \num{0.547} & \num{0.051}\\
Traditional search & \num{-0.300} & \num{0.372}\\
\bottomrule
\end{tabular}
\label{si-tab:attrition}
\end{table}

\newpage

\subsection{Balance check}\label{si:cov-balance}
We checked the balance of all pre-registered pre-treatment covariates and report the results in Table \ref{si-tab:balance-check}. We calculated the mean in experiment groups for every value of the categorical covariates, (i.e. the proportion of participants of that value in a group), and ran a joint test of whether the means are equal across groups. We report both unadjusted p-values and FDR-adjusted p-values. 47 out of 51 categorical levels have unadjusted and adjusted p-values greater than 0.05, and the imbalance observed in the rest is likely due to randomness: these subgroups have small sample sizes, so simple randomization is likely to cause the slight imbalance within these groups. 
\begin{landscape}
    \centering
\begin{longtable}{llcccccccc}
    \caption{Pre-treatment Covariates Balance Check}
    \\ 
    \hline
    Covariate & & GenAI & NonAI & Explanation & Reference & Highlighting & Feedback & p & adj.p\\
    \hline
    \endfirsthead
    \hline
    Covariate & & GenAI & NonAI & Explanation & Reference & Highlighting & Feedback & p & adj. p\\
    \hline
    \endhead
    Age & 18 - 24 & 13.51 & 11.64 & 13.42 & 13.50 & 12.36 & 12.72 & 0.83 & 0.98\\
             & 25 - 34 & 26.41 & 24.11 & 26.85 & 22.93 & 23.50 & 23.45 & 0.30 & 0.93\\
             & 35 - 44 & 20.88 & 21.50 & 20.20 & 22.29 & 23.87 & 21.59 & 0.57 & 0.98\\
             & 45 - 54 & 14.13 & 19.12 & 17.24 & 17.83 & 16.40 & 19.72 & 0.04* & 0.58\\
             & 55 - 64 & 15.97 & 14.25 & 13.79 & 14.90 & 15.54 & 12.95 & 0.52 & 0.95\\
             & 65 or older & 9.09 & 9.38 & 8.50 & 8.54 & 8.32 & 9.57 & 0.93 & 0.99\\
            \hline
            Gender & Female & 49.63 & 51.54 & 47.04 & 45.86 & 51.53 & 48.19 & 0.11 & 0.91\\
             & Male & 48.40 & 45.96 & 50.49 & 51.46 & 46.51 & 49.24 & 0.18 & 0.91\\
             & Non-binary/third gender & 1.84 & 2.14 & 2.34 & 2.29 & 1.47 & 2.57 & 0.68 & 0.98\\
             & Prefer not to answer & 0.12 & 0.36 & 0.12 & 0.38 & 0.49 & 0.00 & 0.29 & 0.93\\
            \hline
            Ethnicity & White & 73.71 & 74.94 & 72.66 & 75.03 & 76.25 & 71.41 & 0.24 & 0.93\\
             & Black & 13.14 & 14.13 & 14.66 & 11.34 & 12.85 & 14.35 & 0.38 & 0.93\\
             & Asian & 7.49 & 8.08 & 7.88 & 7.52 & 7.59 & 8.52 & 0.97 & 0.99\\
             & Hispanic/Latino& 12.53 & 11.40 & 11.21 & 12.61 & 12.48 & 13.89 & 0.60 & 0.98\\
             & Native American& 2.70 & 3.21 & 2.22 & 2.68 & 1.96 & 1.63 & 0.32 & 0.93\\
             & Pacific Islander& 0.49 & 0.59 & 0.49 & 0.38 & 0.24 & 0.35 & 0.91 & 0.99\\
             & Others & 0.86 & 1.07 & 0.86 & 1.66 & 1.35 & 1.40 & 0.62 & 0.98\\
            \hline
            Education & Less than high school & 0.37 & 1.07 & 1.11 & 0.76 & 0.37 & 0.35 & 0.16 & 0.91\\
             & High school graduate & 12.90 & 12.11 & 10.84 & 14.65 & 11.51 & 11.20 & 0.20 & 0.91\\
             & Some college & 19.90 & 21.50 & 23.15 & 19.11 & 20.44 & 20.42 & 0.43 & 0.93\\
             & 2 year degree & 9.83 & 9.14 & 8.37 & 10.45 & 9.91 & 9.33 & 0.79 & 0.98\\
             & 4 year degree & 37.96 & 39.43 & 39.29 & 38.60 & 41.37 & 42.12 & 0.48 & 0.93\\
             & Professional degree & 16.34 & 14.61 & 14.90 & 14.78 & 15.06 & 14.47 & 0.92 & 0.99\\
             & Doctorate & 2.70 & 2.14 & 2.34 & 1.66 & 1.35 & 2.10 & 0.45 & 0.93\\
            \hline
             Political leaning & Strongly republican & 4.05 & 3.92 & 5.42 & 5.99 & 3.55 & 4.20 & 0.13 & 0.91\\
             & Republican & 8.72 & 8.55 & 7.64 & 10.96 & 8.94 & 9.33 & 0.32 & 0.93\\
             & Lean republican & 9.58 & 10.57 & 9.73 & 10.06 & 11.26 & 9.68 & 0.86 & 0.98\\
             & Neutral & 17.69 & 17.22 & 20.07 & 18.85 & 19.34 & 18.55 & 0.70 & 0.98\\
             & Lean democratic & 15.72 & 17.58 & 16.50 & 14.78 & 16.40 & 17.50 & 0.64 & 0.98\\
             & Democratic & 22.11 & 23.28 & 21.31 & 21.15 & 20.81 & 20.65 & 0.79 & 0.98\\
             & Strongly democratic & 22.11 & 18.88 & 19.33 & 18.22 & 19.71 & 20.07 & 0.47 & 0.93\\
             \hline
            Career & Public services & 5.65 & 7.24 & 7.39 & 5.73 & 5.26 & 6.77 & 0.34 & 0.93\\
             & Business\&Finance & 9.83 & 10.21 & 11.21 & 13.25 & 9.91 & 9.92 & 0.19 & 0.91\\
             & Education & 8.97 & 7.60 & 9.48 & 8.79 & 9.42 & 7.70 & 0.61 & 0.98\\
             & Technology & 16.34 & 14.49 & 16.26 & 16.82 & 17.01 & 15.64 & 0.76 & 0.98\\
             & Healthcare & 11.43 & 9.50 & 6.90 & 8.15 & 9.91 & 9.80 & 0.04* & 0.58\\
             & Arts\&Media & 9.71 & 9.26 & 7.51 & 11.21 & 9.91 & 8.52 & 0.18 & 0.91\\
             & Manufacturing & 7.86 & 6.89 & 7.02 & 7.77 & 6.36 & 7.58 & 0.84 & 0.98\\
             & Others & 36.24 & 41.21 & 42.12 & 35.80 & 39.41 & 39.91 & 0.05* & 0.58\\
            \hline
            GenAI& Never & 14.99 & 15.91 & 15.39 & 14.27 & 13.83 & 15.64 & 0.83 & 0.98\\
            usage & Less than once a week & 33.29 & 32.07 & 32.27 & 32.36 & 33.17 & 32.67 & 0.99 & 0.99\\
             & Once a week & 13.76 & 13.66 & 15.02 & 13.76 & 13.95 & 12.72 & 0.86 & 0.98\\
             & 2-3 days a week & 18.55 & 16.75 & 18.35 & 20.00 & 20.44 & 18.67 & 0.46 & 0.93\\
             & 4-6 days a week & 9.46 & 11.52 & 8.87 & 10.70 & 8.81 & 10.62 & 0.34 & 0.93\\
             & Daily & 9.95 & 10.10 & 10.10 & 8.92 & 9.79 & 9.68 & 0.97 & 0.99\\
            \hline
            Generative search& Never & 30.10 & 31.83 & 32.39 & 33.63 & 31.70 & 33.37 & 0.69 & 0.98\\
            usage & Less than once a week & 28.87 & 25.89 & 25.62 & 27.64 & 24.48 & 26.37 & 0.41 & 0.93\\
             & Once a week & 7.99 & 10.45 & 9.61 & 7.64 & 10.89 & 7.12 & 0.03* & 0.58\\
             & 2-3 days a week & 12.53 & 12.47 & 13.42 & 14.65 & 15.18 & 13.89 & 0.52 & 0.95\\
             & 4-6 days a week & 8.48 & 8.31 & 8.37 & 6.75 & 6.85 & 9.22 & 0.37 & 0.93\\
             & Daily & 12.04 & 11.05 & 10.59 & 9.68 & 10.89 & 10.04 & 0.71 & 0.98\\
            \hline
    \end{longtable}
    \label{si-tab:balance-check}
\end{landscape}

\newpage

\section{Analysis Strategy}\label{si:models}
We pre-registered our primary analysis. The main outcome variables are the average of the 5-item trust rating (referred to as ``5-item trust'') and the willingness to share rating. The control variables include age, gender, ethnicity, education level, career industry, political leaning, the frequency of using generative AI and the frequency of using generative search engines. When presenting multiple testing results, we included both original and FDR-adjusted p-values. Error bars are 95\% confidence intervals.

We use model \ref{si-spec:ate} to analyze between-subjects treatment effects, where $t$ indexes the treatment, $\eta_q$ represents search task fixed effects, and $X_i$ is a vector of pre-treatment individual-level covariates. Standard errors are heteroskedasticity-robust. Our parameters of interest are average treatment effects $\beta_t$s. 
\begin{equation}\label{si-spec:ate}
    Y_{iq} = \alpha_0 + \sum_{t}\beta_t(\text{Treatment}_{it})+\gamma{X_i}+\eta_{q}+\epsilon_{iq}
\end{equation}

We use a two-way fixed effects model \ref{si-spec:within} to analyze within-subjects treatment effects in reference, uncertainty highlighting, and social feedback treatment groups. $\eta_q$ is the search task fixed effect, and $\lambda_i$ is the individual fixed effect. Standard errors are clustered at both individual and search task levels. We are interested in the within-subjects treatment effect $\beta$. 
\begin{equation}\label{si-spec:within}
    Y_{iq} = \beta(\text{Variation}_{iq}) +\lambda_{i}+\eta_{q}+\epsilon_{iq}
\end{equation}

To analyze heterogeneous treatment effects, we fit models \ref{si-spec:ate} and \ref{si-spec:within} in subgroups to estimate conditional average treatment effects. 

We use ANOVA to test trust heterogeneity by search task ($H_0: \mu_1 = \mu_2= ... = \mu_9$, i.e. the mean are equal across search tasks). 

As post hoc analysis, we use a random effects model \ref{si-spec:re} to test outcome and treatment effect heterogeneity by search task. We consider random intercepts (difference in outcomes due to search tasks) and random slopes (difference in treatment effects due to search tasks). Our parameters of interest are the variability in the outcome across search tasks $\sigma^2_{b_q}$, and variability in treatment effects across search tasks $\sigma^2_{\gamma_q}$. 
\begin{equation}\label{si-spec:re}
    Y_{iq} = (b_0 + b_q) + (\gamma_0 + \gamma_q) * Treatment_i + \pi X_i + \epsilon_{iq}
\end{equation}

\newpage

\section{Main Analysis Regression Results}
Table \ref{si-tab:ate} and \ref{si-tab:within} present the regression results of the main between-subjects and within-subjects analyses on the entire sample (n = 4927), following the analysis strategy described in \ref{si:models}. 

\begin{table}[tbhp!]
   \caption{Between-subjects Treatment Effects on 5-item Trust and Willingness to Share}\label{si-tab:ate}
   \centering
   \begin{tabular}{lcc} 
   \midrule\midrule
                  & 5-item Trust  & Willingness to Share\\ 
      \\
      Constant                 & 4.328          & 3.907\\   
      \\
      \emph{Treatment}\\
      Traditional Search       & 0.043$^{*}$    & 0.141$^{***}$\\      
                               & (0.020)        & (0.028)\\ 
      Explanation              & -0.011         & 0.011\\  
                               & (0.020)        & (0.029)\\       
      Reference                 & 0.091$^{***}$  & 0.121$^{***}$\\   
                                & (0.020)        & (0.029)\\    
      Uncertainty Highlighting  & -0.157$^{***}$ & -0.179$^{***}$\\      
                              & (0.020)        & (0.029)\\   
      Social Feedback     & 0.018          & 0.022\\    
                              & (0.020)        & (0.028)\\ 
      \\
      Covariate Controls       & Yes            & Yes\\  
      \midrule
      \emph{Fixed Effects}\\
      search task                    & Yes            & Yes\\  
      \midrule
      \emph{Fit statistics}\\
      R$^2$                    & 0.101          & 0.095\\  
      Within Adjusted R$^2$    & 0.049          & 0.058\\  
      Observations             & 44,343         & 44,343\\  
      \midrule\midrule
      \multicolumn{3}{l}{\emph{Heteroskedasticity-robust standard-errors in parentheses}}\\
      \multicolumn{3}{l}{\emph{Signif. Codes: ***: 0.001, **: 0.01, *: 0.05}}\\
   \end{tabular}
\end{table}

\begin{table}[htbp!]
    \caption{Within-subjects Treatment Effects on 5-item Trust and Willingness to Share}\label{si-tab:within}
    \begin{minipage}{\textwidth}
    \centering
        \begin{tabular}{lccc}
        \midrule\midrule
              & Reference & Highlighting & Feedback\\
            \midrule
            \addlinespace[0.5em]
            \multicolumn{4}{l}{\textit{Outcome: 5-item Trust}}\\
            \midrule \hspace{1em}variation & -0.021 & -0.338*** & -0.213***\\
            \hspace{1em} & (0.014) & (0.041) & (0.040)\\
            \hspace{1em}FE: Individual & Yes & Yes & Yes\\
            \hspace{1em}FE: task & Yes & Yes & Yes\\
            \hspace{1em}Observations & 5495\footnote{\label{lessobs}The within-subjects analysis of reference treatment includes 7 out of 9 search tasks. We discuss reasons and details in \ref{si:q68}.} & 7353 & \vphantom{1} 7713\\
            \addlinespace[0.5em]
            \multicolumn{4}{l}{\textit{Outcome: Willingness to Share}}\\
            \midrule \hspace{1em}variation & -0.008 & -0.490*** & -0.286***\\
            \hspace{1em} & (0.026) & (0.057) & (0.051)\\
            \hspace{1em}FE: Individual & Yes & Yes & Yes\\
            \hspace{1em}FE: task & Yes & Yes & Yes\\
            \hspace{1em}Observations & 5495& 7353 & 7713\\
            \midrule\midrule
            \multicolumn{3}{l}{\emph{Two-way clustered standard-errors in parentheses}}\\
             \multicolumn{3}{l}{\emph{Signif. Codes: ***: 0.001, **: 0.01, *: 0.05}}\\
        \end{tabular}
        \end{minipage}
\end{table}

\newpage

\section{Discussion on Inaccurate Generative Responses}\label{si:q68}
We observed hallucinations when collecting generative search response data. The response to ``Government policy to fight inflation'' displayed fact and context conflicting hallucinations: it included fiscal policies for controlling \textit{deflation} (``increasing government spending and cutting tax rate''), and cited an article on ``7 Ways Governments Fight Deflation.'' The response to ``How often do adults need vaccines'' displayed reference-conflicting hallucinations: the response claimed that a booster dose of either Tdap or Td every 10 years is ``recommended for age 19-26'', while the referenced article recommends a much wider age range. Because these two responses are inaccurate, they don't have the baseline stimuli of the reference treatment (which requires references to be valid). Therefore we excluded these two tasks from the within-subjects analysis of the reference treatment.

\newpage

\section{Invalid Reference Types}
Our within-subjects analysis of the reference treatment is based on ratings of 7 search tasks: a random selection of three cite invalid references, while references in the other responses are valid. In addition, when creating variation stimuli for the 7 tasks, in five of them we replaced one reference link with a \textit{invalid and broken} one, and in the other two we replaced one link with a clickable but \textit{irrelevant} one. We replicated the within-subjects analysis for these two cases separately. Table \ref{si-tab:reftype-within} shows that participants did not react to either type of invalidity. \begin{table}[htbp!]
\caption{Within-subjects Invalid Reference effects, by Invalid Reference Types}
    \begin{minipage}{\textwidth}
    \centering
    \begin{tabular}[t]{lccc}
        \midrule
          & All & Broken & Irrelevant\\
        \midrule
        \addlinespace[0.5em]
        \multicolumn{4}{l}{\textit{Outcome: 5-item Trust}}\\
        \midrule \hspace{1em}Invalid reference & -0.021 & -0.021 & -0.009\\
        \hspace{1em} & (0.014) & (0.017) & (0.066)\\
        \addlinespace[0.5em]
        \hspace{1em}Observations & 5495 & 3925 & \vphantom{1} 1570\\
        \hspace{1em}FE: Individual & Yes & Yes & Yes\\
        \hspace{1em}FE: task & Yes & Yes & Yes\\
        \addlinespace[0.5em]
        \midrule
        \multicolumn{4}{l}{\textit{Outcome: Willingness to Share}}\\
        \midrule \hspace{1em}Invalid reference & -0.008 & -0.024 & 0.042\\
        \hspace{1em} & (0.026) & (0.029) & (0.089)\\
        \addlinespace[0.5em]
        \hspace{1em}Observations & 5495 & 3925 & 1570\\
        \hspace{1em}FE: Individual & Yes & Yes & Yes\\
        \hspace{1em}FE: task & Yes & Yes & Yes\\
        \midrule
        \multicolumn{3}{l}{\emph{Two-way clustered standard-errors in parentheses}}\\
        \multicolumn{3}{l}{\rule{0pt}{1em}* p $<$ 0.05, ** p $<$ 0.01, *** p $<$ 0.001}\\
    \end{tabular}
    \label{si-tab:reftype-within}
    \end{minipage}
\end{table}

\newpage
\section{Clicks and Time Spent Analysis}\label{si:timeclick}

We examined time spent per task and clicks per task as two additional dependent variables. In our experiment, we presented each search task on a separate page, showing each search search query, the corresponding generative response, clickable reference links (where available), and trust rating questions. Time spent per task is the duration of time that participants spent on a search task page, and clicks per task is the number of clicks they made on a page. We used these two variables to approximate participants' interactions with a search task page.

Table \ref{si-tab:timeclick-trust} presents the regression results of the time spent per task (in seconds) and the number of clicks per task on the 5-item trust score, using all ratings from participants in the GenAI and the GenAI variant conditions. Table \ref{si-tab:timeclick-ate}, \ref{si-tab:timeclick-wi}, \ref{si-tab:timeclick-reftype} show the results from between-subjects and within-subjects comparisons with these two variables. We noticed that participants spent more time when they were given reference links, and spent less time when provided with uncertainty highlighting or when they were told in the experiment that the response was not created by AI. Moreover, the highlighting treatment group spent much less time on tasks that show high uncertainty highlighting only. 

We also checked the variation of these two variables by order and by search task. Fig. \ref{si-fig:timeclick_task} shows group-wise comparisons of time spent by task, in ascending order of their 5-item trust rating in the GenAI group (i.e., the left is the least trusted and the right is the most trusted in the GenAI group). Participants seemed to spend less time on the top two most trusted search tasks, but in general there is no clear pattern. 

Conversely, Fig.\ref{si-fig:timeclick_pos} displays strong order effects. Regardless of the group that participants were assigned to, the first task took them the longest time, and they clicked the first task page more. The reference treatment group spent more time in the first three tasks compared to all other groups, while the difference later diminished. Similarly, the page clicks difference between reference and GenAI groups, i.e. induced by reference treatment, decreases in order. Furthermore, participants given uncertainty highlighting spent less time compared to the other groups over the entire experiment. These results validate our decision to randomize the order of search tasks, which controlled for order effects in our experimental results.

\subsection{Discussion of Reference Clicking and Checking Behaviors}\label{si:firstinvalid}
Participants spent more time on the first task when provided with references. They also clicked the first task page more. These observations suggest that reference treatment participants paid more attention to and spent more effort investigating the first search task response. We considered this possibility and hypothesized that they would be more likely to notice if the first task contains invalid references.  We therefore tested whether the first task containing invalid links predicted participants' trust in that task and their overall trust level (carry-over effects), but the results were not significant (see Table \ref{si-tab:first_invalid}). In summary, we did not find supporting evidence of participants noticing invalid references even when they spent more time and more effort.

\begin{table}[htbp!]
   \caption{Regression Analysis of Time Spent and Clicks on the 5-item Trust Score}
   \centering
   \begin{tabular}{lcc}
      \midrule
                               & Times Spent (sec.) & Clicks \\  
     \\
      (Intercept) & \num{56.573}*** & \num{7.997}*** \\
      & (\num{1.148})   & (\num{0.101})  \\ 
      5-item trust        & \num{-1.139}*** & \num{0.107}*** \\
    & (\num{0.234})   & (\num{0.022})  \\
      \midrule
      \emph{Fit statistics}\\
      R2          & \num{0.001}     & \num{0.001}    \\
R2 Adj.     & \num{0.000}     & \num{0.001}    \\
RMSE        & \num{63.70}     & \num{5.37}     \\
      Observations             & 36,765     & 36,765 \\
      \midrule \midrule
      \multicolumn{3}{l}{\emph{Heteroskedasticity-robust standard-errors in parentheses}}\\
      \multicolumn{3}{l}{\emph{Signif. Codes: ***: 0.001, **: 0.01, *: 0.05}}\\
   \end{tabular}
   \label{si-tab:timeclick-trust}
\end{table}

\begin{table}[htbp!]
   \caption{Between-subjects Effects on Time Spent and Clicks}
   \centering
   \begin{tabular}{lcc}
      \midrule
                               & Times Spent (sec.) & Clicks \\  
     \\
      \emph{Treatment}\\
      Traditional search       & -2.517$^{*}$   & -0.566$^{***}$\\     
                               & (1.160)        & (0.085)\\  
      Explanation            & -1.347         & -0.444$^{***}$\\    
                              & (1.042)        & (0.090)\\   
      Reference               & 5.943$^{***}$  & 0.327$^{**}$\\   
                              & (1.172)        & (0.101)\\  
      Uncertainty Highlighting  & -5.074$^{***}$ & -0.410$^{***}$\\  
                               & (0.984)        & (0.090)\\    
      Social Feedback          & -1.657         & -0.434$^{***}$\\  
                                & (1.003)        & (0.090)\\ 
        \\
      Covariate Controls       & Yes            & Yes\\  
      \midrule
      \emph{Fixed Effects}\\
      query                    & Yes              & Yes\\  
      \midrule
      \emph{Fit statistics}\\
      R$^2$                    & 0.032            & 0.015\\  
      Within Adjusted R$^2$    & 0.025            & 0.014\\  
      Observations             & 44,343           & 44,343\\  
      \midrule \midrule
      \multicolumn{3}{l}{\emph{Heteroskedasticity-robust standard-errors in parentheses}}\\
      \multicolumn{3}{l}{\emph{Signif. Codes: ***: 0.001, **: 0.01, *: 0.05}}\\
   \end{tabular}
   \label{si-tab:timeclick-ate}
\end{table}

\begin{table}[htbp!]
\caption{Within-subjects Effects on Time Spent and Clicks}
\centering
    \begin{tabular}[t]{lccc}
    \midrule
      & Reference & Highlighting & Feedback\\
    \midrule
    \addlinespace[0.5em]
    \multicolumn{4}{l}{\textit{Outcome: Time spent}}\\
    \midrule 
    \hspace{1em}variation & 1.187    & -2.747*  & 0.388  \\
    \addlinespace[0.5em]
    \hspace{1em} & (2.516)& (0.862)  & (1.499)\\
    \hspace{1em}Observations & 5495 & 7353 & \vphantom{1} 7713\\
    \hspace{1em}FE: Individual & Yes & Yes & Yes\\
    \hspace{1em}FE: task & Yes & Yes & Yes\\
    \addlinespace[0.5em]
    \multicolumn{4}{l}{\textit{Outcome: Clicks}}\\
    \midrule \hspace{1em}variation & 0.066   & 0.058   & 0.078 \\
    \hspace{1em} & (0.084)  & (0.072)  & (0.040)\\
    \addlinespace[0.5em]
    \hspace{1em}Observations & 5495 & 7353 & \vphantom{1} 7713\\
    \hspace{1em}FE: Individual & Yes & Yes & Yes\\
    \hspace{1em}FE: task & Yes & Yes & Yes\\
    \bottomrule
    \multicolumn{3}{l}{\emph{Two-way clustered standard-errors in parentheses}}\\
    \multicolumn{3}{l}{\rule{0pt}{1em}* p $<$ 0.05, ** p $<$ 0.01, *** p $<$ 0.001}\\
    \end{tabular}
    \label{si-tab:timeclick-wi}
\end{table}

\begin{table}[htbp!]
\caption{Effects of Invalid Reference on Time Spent and Clicks}
\begin{minipage}{\textwidth}
\centering
    \begin{tabular}[t]{lccc}
    \midrule
          & All & Broken & Irrelevant\\
        \midrule
        \addlinespace[0.5em]
        \multicolumn{4}{l}{\textit{Outcome: Time spent (in sd.)}}\\
        \midrule 
        \hspace{1em}Invalid reference & 1.187& 3.779  & -6.954  \\
        \hspace{1em}& (2.516)   & (2.298)   & (3.607) \\
        \addlinespace[0.5em]
        \hspace{1em}Observations & 5495 & 3925 & 1570\\
        \hspace{1em}FE: Individual & Yes & Yes & Yes\\
        \hspace{1em}FE: task & Yes & Yes & Yes\\
        \addlinespace[0.5em]
        \multicolumn{4}{l}{\textit{Outcome: Clicks (in sd.)}}\\
        \midrule 
        \hspace{1em}Invalid reference & 0.066  & -0.052 & 0.257\\
        \hspace{1em} & (0.084)    & (0.067)    & (0.199) \\
        \addlinespace[0.5em]
        \hspace{1em}Observations & 5495 & 3925 & 1570\\
        \hspace{1em}FE: Individual & Yes & Yes & Yes\\
        \hspace{1em}FE: task & Yes & Yes & Yes\\
        \midrule
        \multicolumn{3}{l}{\emph{Two-way clustered standard-errors in parentheses}}\\
        \multicolumn{3}{l}{\rule{0pt}{1em}* p $<$ 0.05, ** p $<$ 0.01, *** p $<$ 0.001}\\
    \end{tabular}
    \label{si-tab:timeclick-reftype}
\end{minipage}
\end{table}

\begin{figure}[htbp!]
    \caption{Between-subjects Effects on Time Spent and Clicks, Grouped by Task}
    \centering
    \begin{subfigure}[t]{\textwidth}
    \centering
    \includegraphics[width=\textwidth]{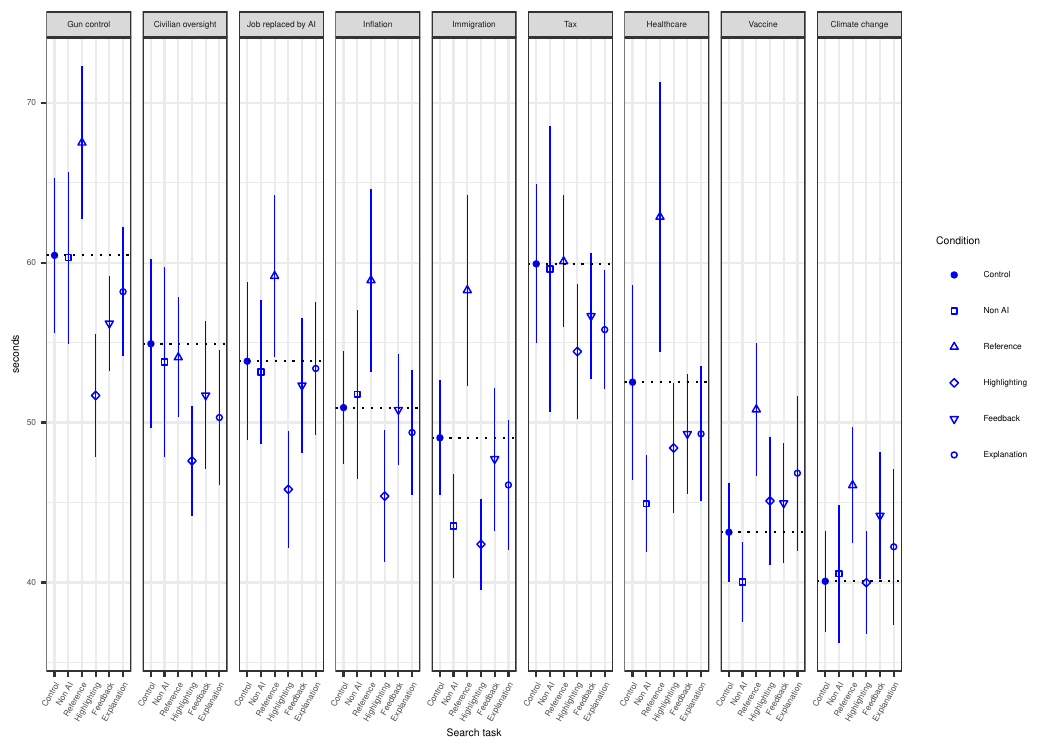} 
    \caption{Time Spent} \label{si-fig:time_task}
    \end{subfigure}
    \begin{subfigure}[t]{\textwidth}
    \centering
    \includegraphics[width=\textwidth]{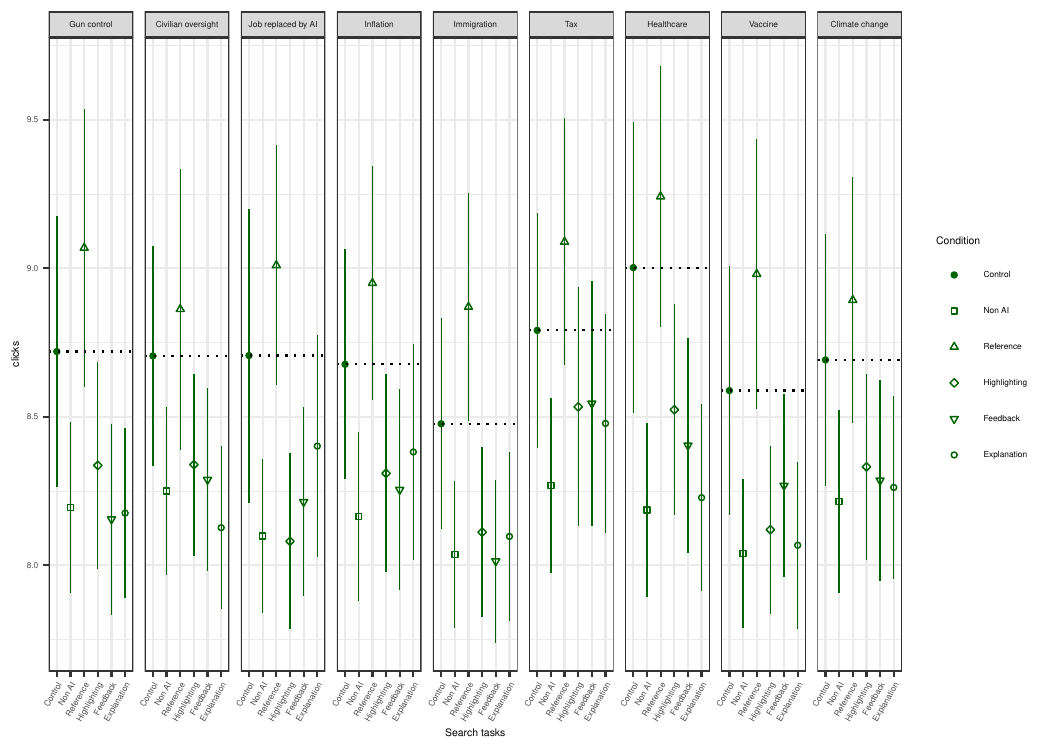} 
    \caption{Clicks} \label{si-fig:click_task}
    \end{subfigure}
\end{figure}\label{si-fig:timeclick_task}

\begin{figure}[htbp!]
    \caption{Between-subjects Effects on Time Spent and Clicks, Grouped by Task Order}
    \centering
    \begin{subfigure}[t]{\textwidth}
    \centering
    \includegraphics[width=\textwidth]{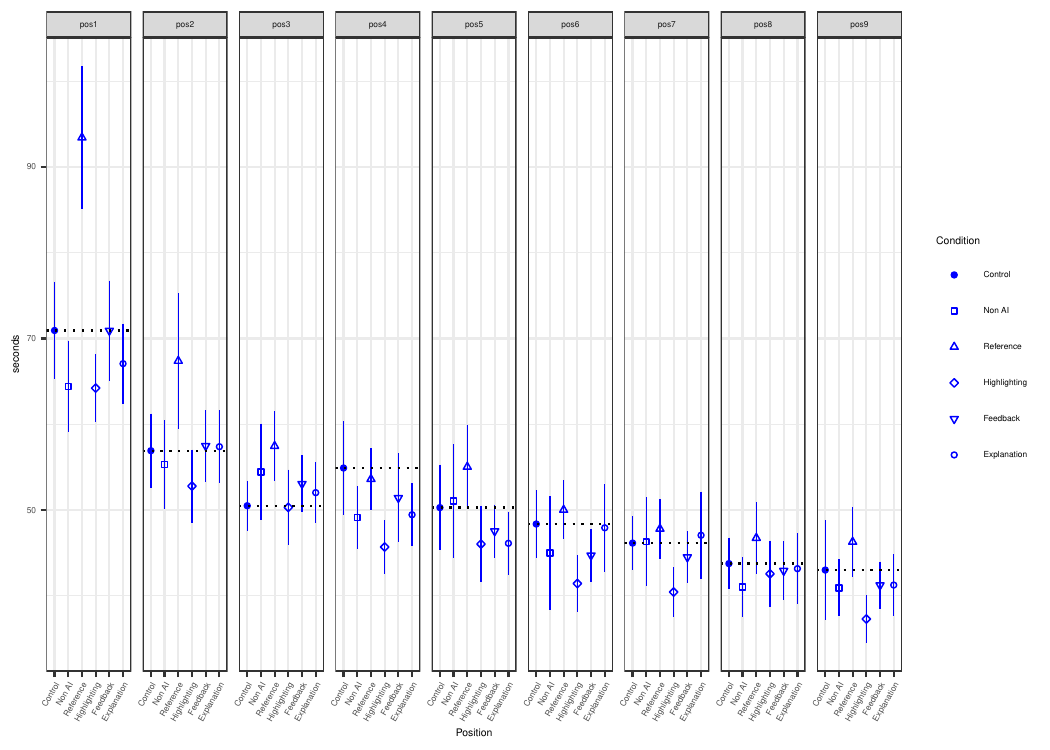} 
    \caption{Time Spent} \label{si-fig:time_pos}
    \end{subfigure}
    \begin{subfigure}[t]{\textwidth}
    \centering
    \includegraphics[width=\textwidth]{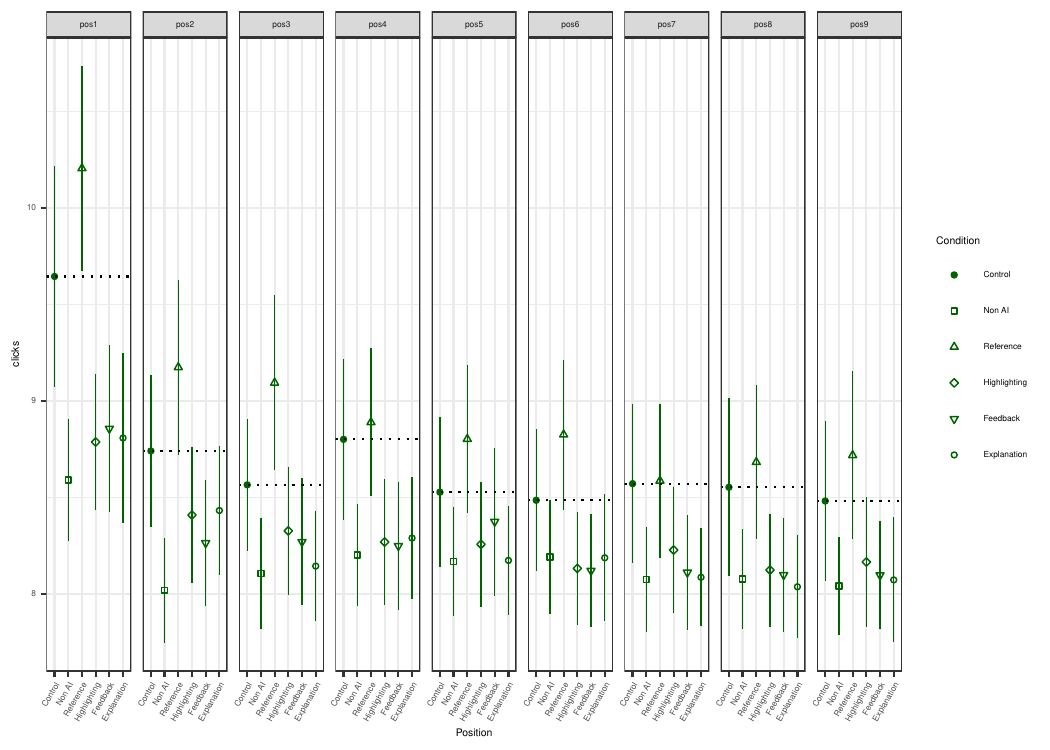} 
    \caption{Clicks} \label{si-fig:click_pos}
    \end{subfigure}
\end{figure}\label{si-fig:timeclick_pos}

\begin{table}
\caption{Predicting Trust Measures With Whether the First Task Contained Invalid References}
\centering
\begin{minipage}{\textwidth}
\makebox[\linewidth][c]{%
\begin{tabular}[t]{lcccc}
\midrule
  &5-item trust & Willingness to share &5-item trust  & Willingness to share \\
\midrule
first\_invalid & -0.028(0.089)&0.076(0.130)&-0.024(0.031)&0.083(0.043)\\
p-value & 0.754 & 0.557 &0.440 &0.055\\
\addlinespace[0.5em]
Covariate Controls\footnote{Include the eight pre-treatment covariates described in \ref{si:models} and a categorical variable indexing search tasks}  & Yes & Yes& Yes  & Yes\\  
\midrule
Treatment group & Reference only & Reference only & Reference only & Reference only\\
Task & first task only & first task only & all & all\\
Observations & 785 & 785  & 7065& 7065\\
\midrule
R2 & 0.129 & 0.111 & 0.106 & 0.093\\
R2 Adj. & 0.092 & 0.074 0.101 &	0.089\\
RMSE& 1.01 & 1.57 & 1.18 & 1.69\\
\bottomrule
\multicolumn{3}{l}{\emph{Heteroskedasticity-robust standard-errors in parentheses}}\\
\end{tabular}%
}
\label{si-tab:first_invalid}
\end{minipage}
\end{table}

\newpage
\section{Treatment Effects Heterogeneity}
\begin{figure}
    \centering
    \includegraphics[width=1.2\linewidth]{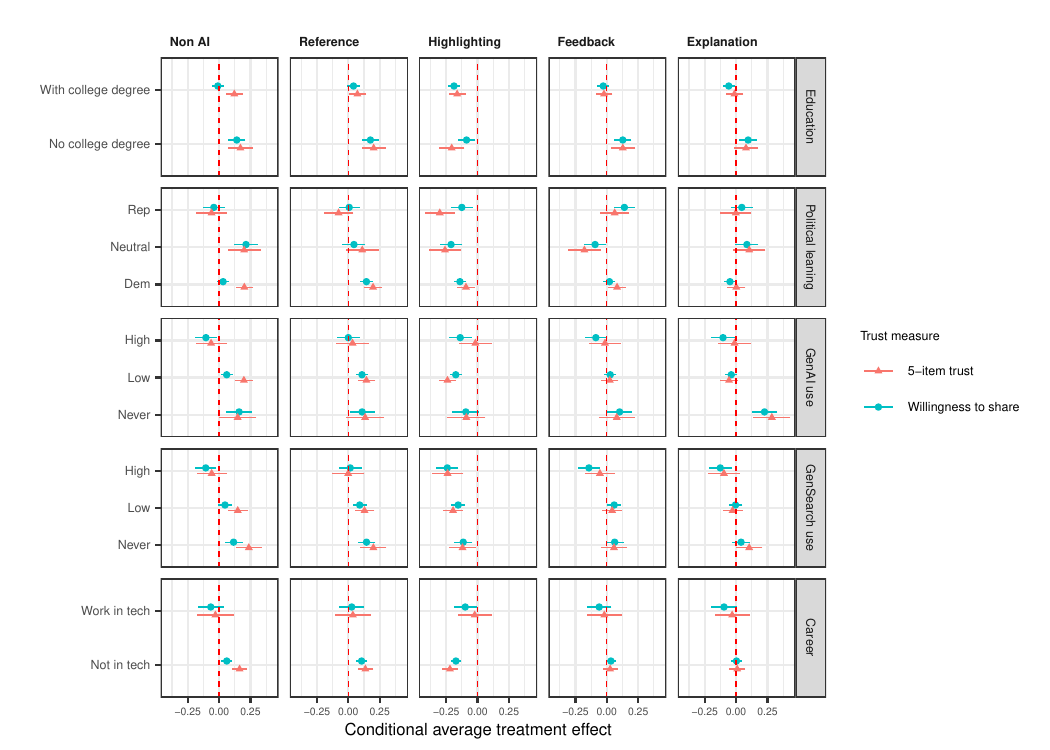}
    \caption{Conditional Average Treatment Effects}
    \label{si-fig:hte5}
\end{figure}

\begin{figure}[htbp!]
    \centering
    \begin{subfigure}[b]{1.2\textwidth}
        \centering
        \includegraphics[width=1\linewidth]{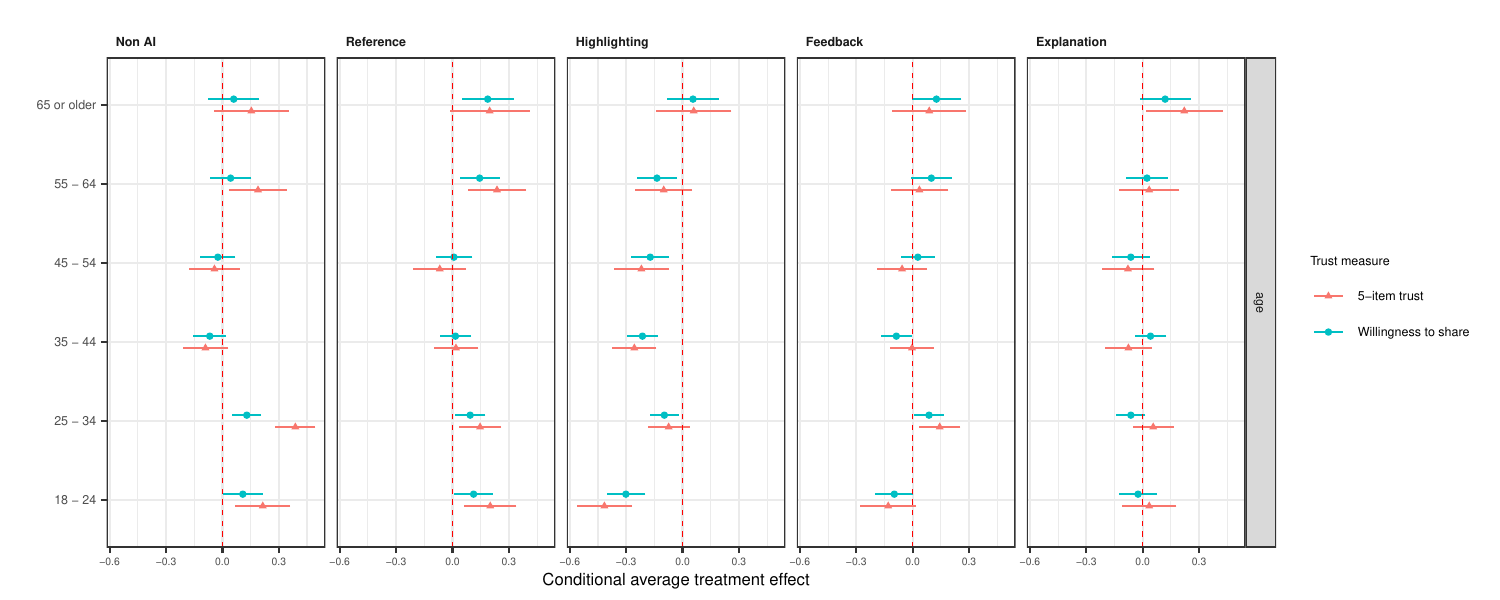}
        \caption{Age}
    \end{subfigure}
    \vskip\baselineskip
    \begin{subfigure}[b]{1.2\textwidth}
        \centering
        \includegraphics[width=1\textwidth]{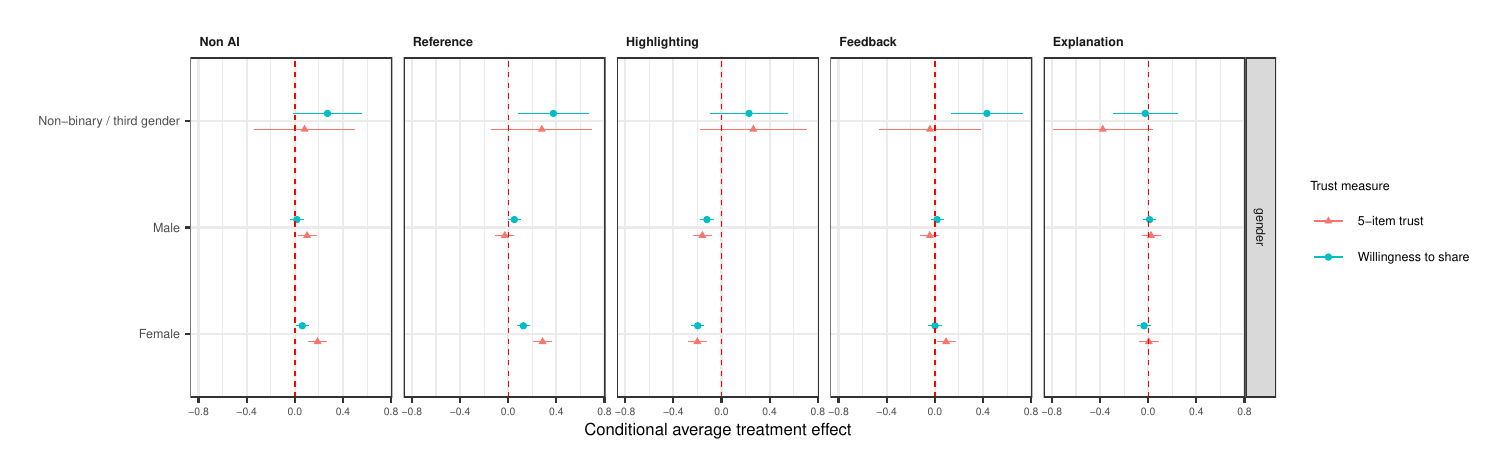}
        \caption{Gender}
    \end{subfigure}
    \vskip\baselineskip
    \begin{subfigure}[b]{1.2\textwidth}
        \centering
        \includegraphics[width=1\linewidth]{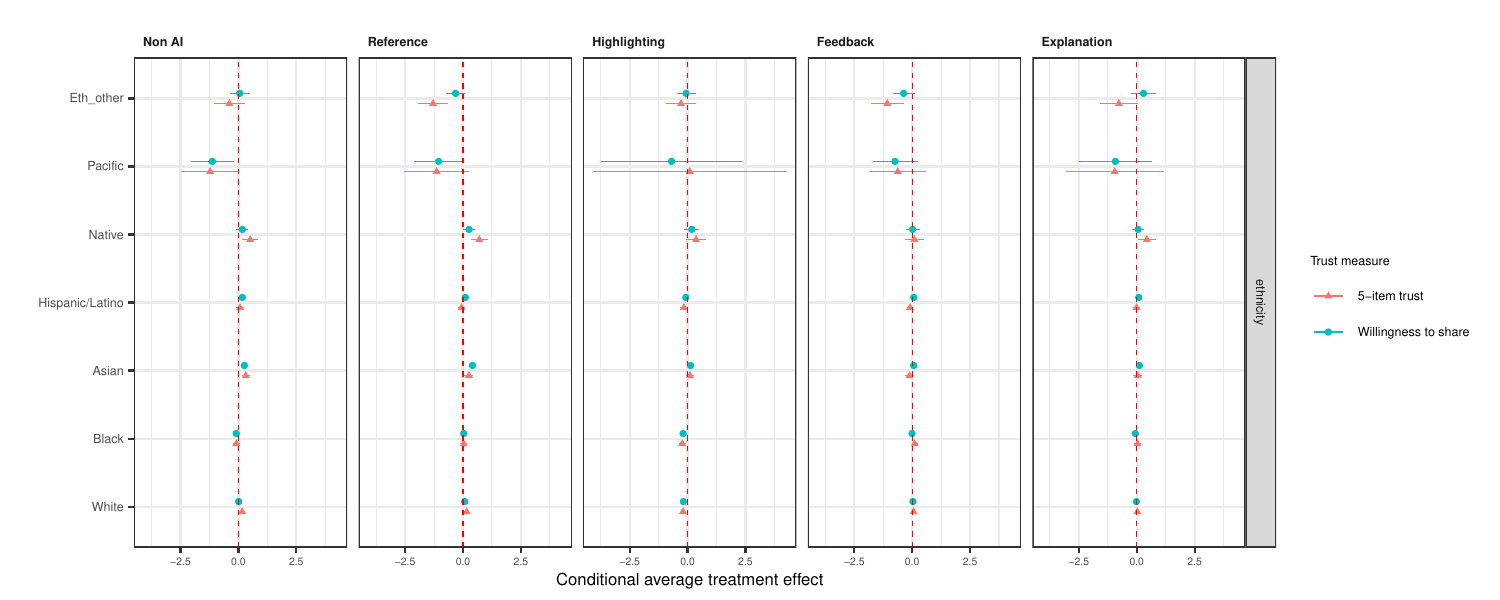}
        \caption{Ethnicity}
    \end{subfigure}
    \caption{Conditional Average Treatment Effects}\label{si-fig:cate-other}
\end{figure}

\begin{figure}
    \centering
    \includegraphics[width=1.2\linewidth]{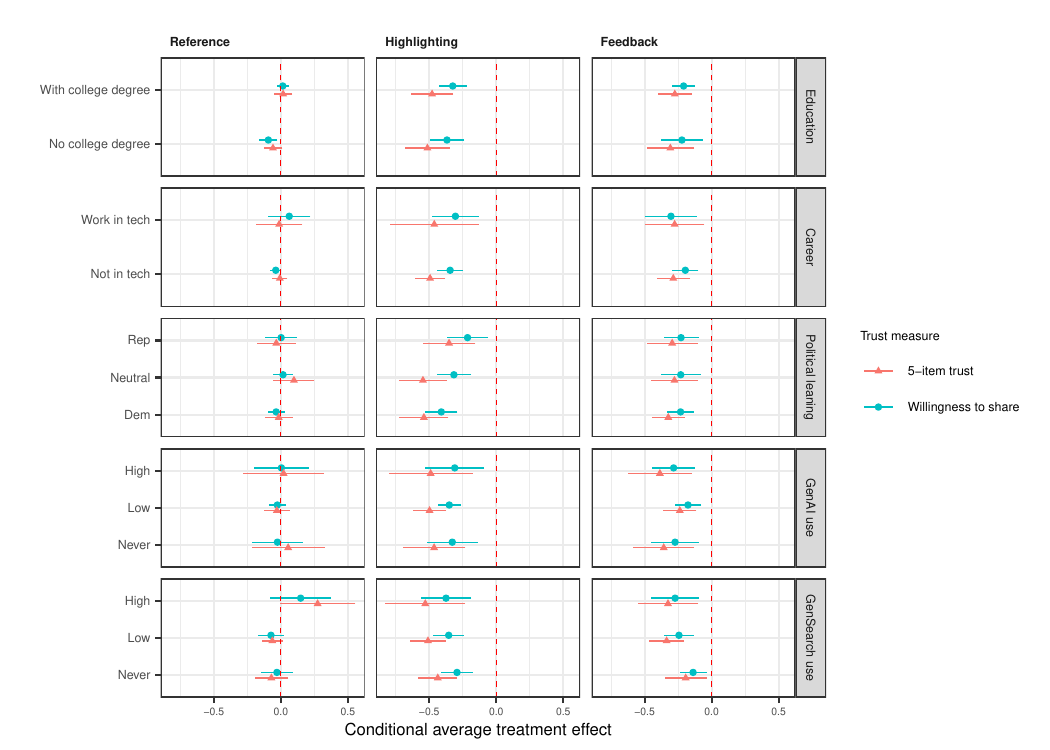}
    \caption{Within-subjects Effects in Subgroups}
    \label{si-fig:wi-hte}
\end{figure}

\begin{figure}[htbp!]
    \centering
    \begin{subfigure}[b]{1.2\textwidth}
        \centering
        \includegraphics[width=1\linewidth]{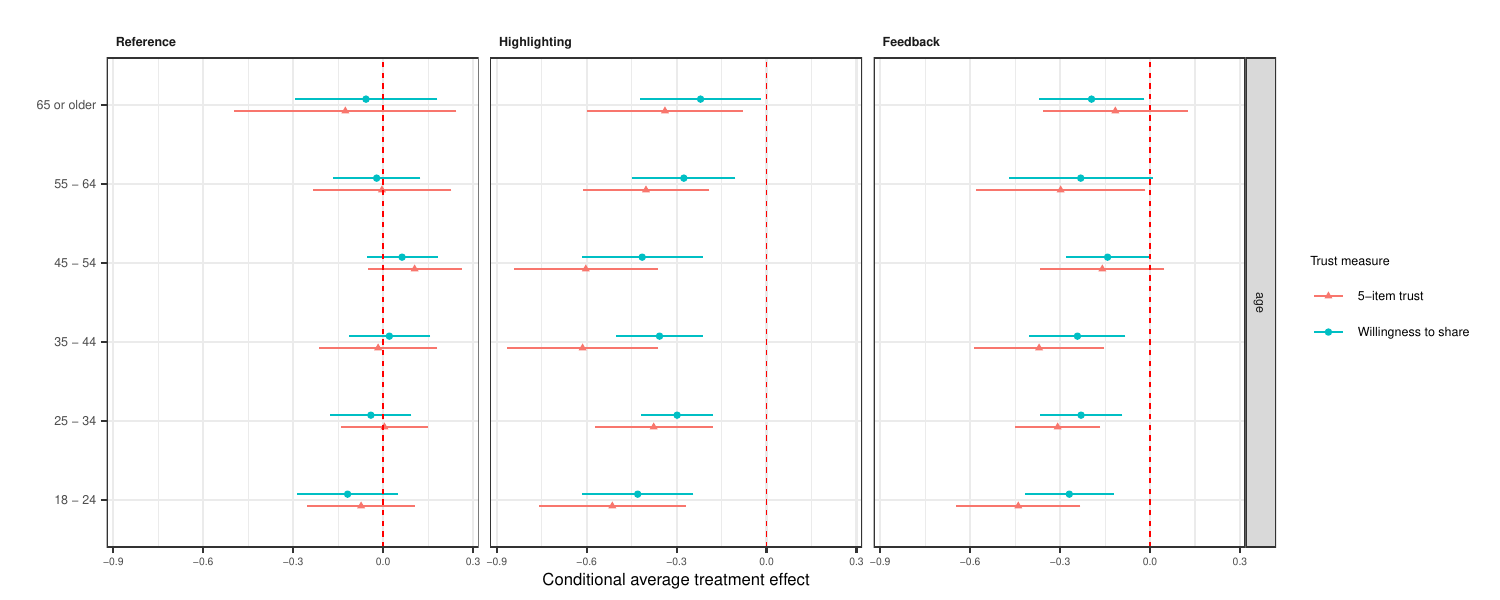}
        \caption{Age}
    \end{subfigure}
    \vskip\baselineskip
    \begin{subfigure}[b]{1.2\textwidth}
        \centering
        \includegraphics[width=1\textwidth]{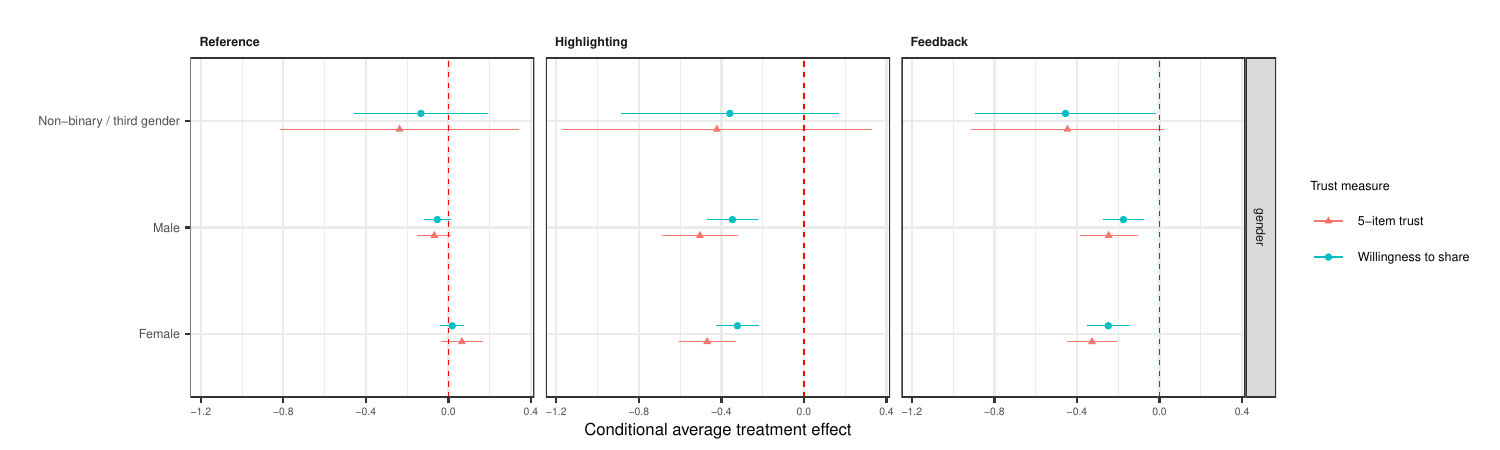}
        \caption{Gender}
    \end{subfigure}
    \vskip\baselineskip
    \begin{subfigure}[b]{1.2\textwidth}
        \centering
        \includegraphics[width=1\linewidth]{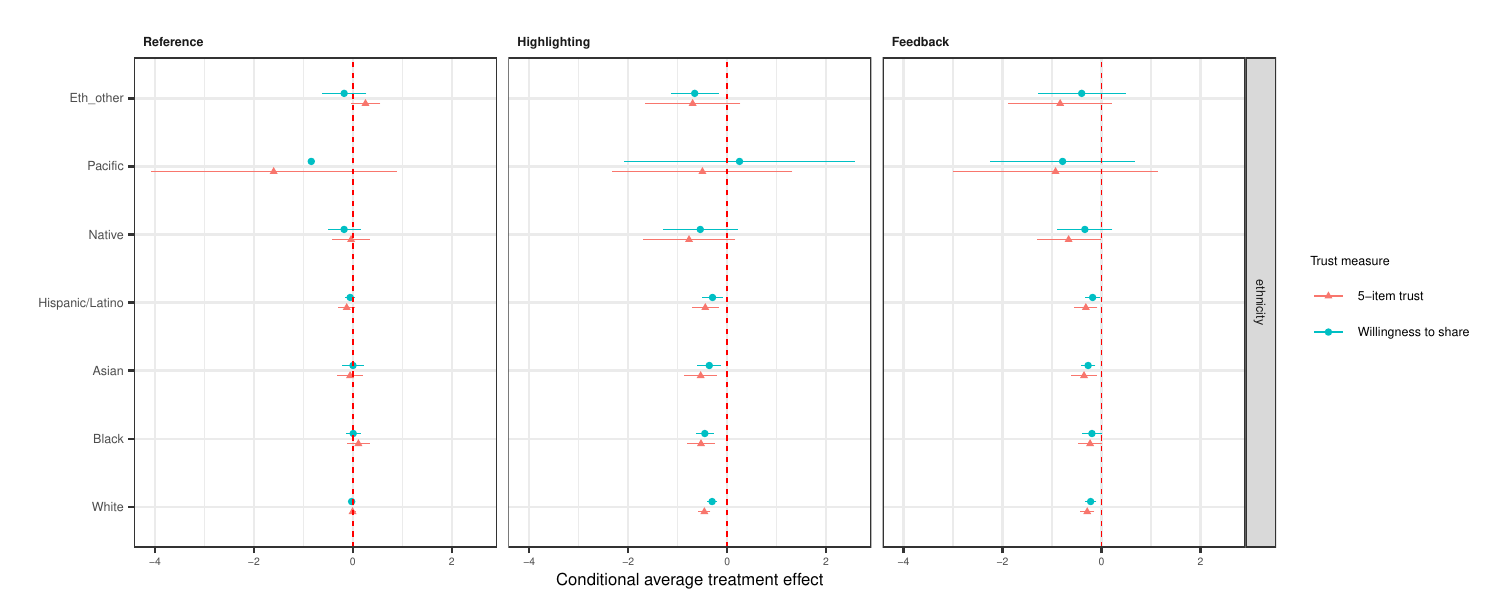}
        \caption{Ethnicity}
    \end{subfigure}
    \caption{Within-subjects Effects in Subgroups}\label{si-fig:wicate-other}
\end{figure}

In figure \ref{si-fig:hte5}, \ref{si-fig:cate-other}, \ref{si-fig:wi-hte}, \ref{si-fig:wicate-other} we present all between-subjects and within-subjects effect heterogeneity results, some of which are provided in the main manuscript as well. Those subgroups with wide confidence intervals are the groups with small group sizes (e.g. binary/third gender, pacific islanders). In general we do not observe clear patterns of heterogeneity in age, gender, and ethnicity, except in Fig. \ref{si-fig:wicate-other} we notice a trend of younger participants lower their trust more if the social feedback is negative, compared to a positive one. 

\newpage

\newpage
\section{Test of Search Task Heterogeneity}\label{si:task-hetero}
We conducted an ANOVA test on the null hypothesis of equal means across search task $ H_0:\mu_1 = \mu_2 = ... = \mu_9)$ and found trust heterogeneity by search task in every group. Results are presented in Table \ref{si-tab:equal-task-mean} with FDR-adjusted p-values.

In a post hoc analysis, we used model \ref{si-spec:re} to further verify outcome and treatment effects heterogeneity by search task. The test of $H_0: \sigma^2_{b_q} = 0 \text{ vs. } H_1: \sigma^2_{b_q} > 0 $ showed heterogeneity in the two trust measures. When testing for random slopes  $H_0: \sigma^2_{\gamma_q} = 0 \text{ vs. } H_1: \sigma^2_{\gamma_q} > 0 $, we found treatment effect heterogeneity on the 5-item trust rating exists for traditional search, reference, uncertainty hightlighting, and treatment effect heterogeneity on the willingness to share for uncertainty highlighting. The results are reported in Table \ref{si-tab:trust_intercept}, \ref{si-tab:share_intercept}, \ref{si-tab:trust_slope}, \ref{si-tab:share_slope}.  

\newpage
\begin{table}[htbp!]
\caption{Joint Tests of Equal Means Across Search Tasks}
    \centering
    \begin{tabular}{l|llcc}
    \hline
        Trust measure & Sample & F & p-value & adj.p\\
        \hline
        5-item Trust& GenAI & 60.97 & 0 & 0\\
        & Traditional search & 78.83 & 0 & 0\\
        & Reference & 40.92 & 0 & 0\\
        & Highlighting & 34.98 & 0 & 0\\
        & Explanation & 57.95 & 0 & 0\\
        & Feedback & 59.68 & 0 & 0\\
        \hline
        \addlinespace[1em]
        \hline
        Trust measure & Sample & F & p-value & adj.p\\
        \hline
        Willingness to share & GenAI & 41.19 & 0 & 0\\
         & Traditional search & 55.62 & 0 & 0\\
         & Reference & 29.70 & 0 & 0\\
         & Highlighting & 27.43 & 0 & 0\\
         & Explanation & 40.75 & 0 & 0\\
         & Feedback & 40.26 & 0 & 0\\
        \hline
\end{tabular}
\label{si-tab:equal-task-mean}
\end{table}

\begin{table}[htbp!]
\caption{Testing Search Task Heterogeneity in 5-item Trust Using Random-effects Models}
\centering
\makebox[\linewidth][c]{%
\begin{tabular}[t]{lrrrrrrrrr}
    \hline
     & npar & AIC & BIC & logLik & deviance & $\chi^2$ & Df & $P(>\chi^2)$ & adj.p\\
     \hline
     \textit{All treatments included} & & & & & & & & &\\
    null model & 30 & 144956 & 145217 & -72448 & 144896  &  &  &  & \\
    + random intercepts& 31 & 142481 & 142751 & -71210 & 142419 & 2477 &  1   &0.00&0.00\\
    \hline
    \textit{Treatment: Traditional search} & & & & & & & & &\\
    null model & 26 & 49246 & 49443 & -24597 & 49194 &  &  &  & \\
    + random intercepts & 27 & 48195 & 48400 & -24070 & 48141 & 1053 & 1 & 0.00 & 0.00\\
    \hline
    \textit{Treatment: Explanation} & & & & & & & & &\\
    null model & 26 & 48120 & 48317 & -24034 & 48068 &  &  &  & \\
    + random intercepts & 27 & 47218 & 47423 & -23582 & 47164 & 903 & 1 & 0.00 & 0.00\\
    \hline
    \textit{Treatment: Reference} & & & & & & & & &\\
    null model & 26 & 46839 & 47036 & -23393 & 46787 &  &  &  & \\
    + random intercepts & 27 & 46075 & 46280 & -23011 & 46021 & 766 & 1 & 0.00 & 0.00\\
    \hline
    \textit{Treatment: Highlighting} & & & & & & & & &\\
    null model & 26 & 47584 & 47781 & -23766 & 47532&  &  &  & \\
    + random intercepts & 27 & 46899 & 47104 & -23422 & 46845 & 687 & 1 & 0.00 & 0.00\\
    \hline
    \textit{Treatment: Feedback} & & & & & & & & &\\
    null model & 26 & 49015 & 49213 & -24481 & 48963 &  &  &  & \\
    + random intercepts & 27 & 48097 & 48303 & -24021 & 48043 & 920 & 1 & 0.00 & 0.00\\
    \hline
    \multicolumn{8}{l}{The null model is model \ref{si-spec:re} with no random effect terms;}\\
    \multicolumn{8}{l}{ the alternative model is model \ref{si-spec:re} with $b_q$ and without $\gamma_q$.}\\
    \end{tabular}%
    }
    \label{si-tab:trust_intercept}
\end{table}

\newpage

\begin{table}[htbp!]
\caption{Testing Search Task Heterogeneity in Willingness to Share Using Random-effects Models}
\centering
\makebox[\linewidth][c]{%
\begin{tabular}[t]{lrrrrrrrrr}
    \hline
     & npar & AIC & BIC & logLik & deviance & $\chi^2$ & Df & $P(>\chi^2)$ & adj.p\\
     \hline
     \textit{All treatments included} & & & & & & & & &\\
    null model &30	&175667	&175928&	-87804&	175607	  &  &  &  & \\
    + random intercepts& 31	&173907&	174177&	-86923&	173845&	1762&	1 &0.00&0.00\\
    \hline
    \textit{Treatment: Traditional search} & & & & & & & & &\\
    null model & 26 & 59216 & 59414 & -29582 & 59164&  &  &  & \\
    + random intercepts & 27 & 58491 & 58697 & -29219 & 58437 & 727 & 1 & 0.00 & 0.00\\
    \hline
    \textit{Treatment: Explanation} & & & & & & & & &\\
    null model & 26 & 58443 & 58640 & -29195 & 58391 &  &  &  & \\
    + random intercepts & 27 & 57822 & 58027 & -28884 & 57768 & 623 & 1 & 0.00& 0.00\\
    \hline
    \textit{Treatment: Reference} & & & & & & & & &\\
    null model & 26 & 57025 & 57222 & -28487 & 56973 &  &  &  & \\
    + random intercepts& 27 & 56501 & 56706 & -28224 & 56447 & 526 & 1 & 0.00 & 0.00\\
    \hline
    \textit{Treatment: Highlighting} & & & & & & & & &\\
    null model & 26 & 58177 & 58374 & -29062 & 58125&  &  &  & \\
    + random intercepts & 27 & 57689 & 57894 & -28818 & 57635 & 489 & 1 & 0.00 & 0.00\\
    \hline
    \textit{Treatment: Feedback} & & & & & & & & &\\
    null model & 26 & 59636 & 59834 & -29792 & 59584 &  &  &  & \\
    + random intercepts & 27 & 59011 & 59217 & -29478 & 58957 & 627 & 1 & 0.00 & 0.00\\
    \hline
    \multicolumn{8}{l}{The null model is model \ref{si-spec:re} with no random effect terms;}\\
    \multicolumn{8}{l}{ the alternative model is model \ref{si-spec:re} with $b_q$ and without $\gamma_q$.}\\
\end{tabular}%
}
\label{si-tab:share_intercept}
\end{table}

\newpage

\begin{table}[htbp!]
\caption{Testing Treatment Effects Heterogeneity on 5-item Trust}
\centering
\makebox[\linewidth][c]{%
\begin{tabular}[t]{lrrrrrrrrr}
    \hline
     & npar & AIC & BIC & logLik & deviance & $\chi^2$ & Df & $P(>\chi^2)$ & adj.p\\
     \hline
     \textit{All treatments included} & & & & & & & & &\\
    null model(random intercept)& 31 & 142481 & 142751 & -71210 & 142419&  &  &  & \\
    + random slopes & 51 & 142409 &142853 & -71154 & 142307 & 112 & 20 & 0.00 & 0.00\\
    \hline
    \textit{Treatment: Traditional search} & & & & & & & & &\\
    null model(random intercept)& 27 & 48195 & 48400 & -24070 & 48141&  &  &  & \\
    + random slopes & 29 & 48190 & 48410 & -24066 & 48132 & 9 & 2 & 0.01 & 0.03\\
    \hline
    \textit{Treatment: Explanation} & & & & & & & & &\\
    null model(random intercept) & 27 & 47218 & 47423 & -23582 & 47164  &  &  &  & \\
    + random slopes & 29 & 47222 & 47442 & -23582 & 47164 & 0 & 2 & 0.95 & 1.00\\
    \hline
    \textit{Treatment: Reference} & & & & & & & & &\\
    null model(random intercept) & 27 & 46075 & 46280 & -23011 & 46021 &  &  &  & \\
    + random slopes& 29 & 46069 & 46289 & -23006 & 46011 & 10 & 2 & 0.01 & 0.03\\
    \hline
    \textit{Treatment: Highlighting} & & & & & & & & &\\
    null model(random intercept) & 27 & 46899 & 47104 & -23422 & 46845 &  &  &  & \\
    + random slopes & 29 & 46869 & 47089 & -23405 & 46811 & 34 & 2 & 0.00 & 0.00\\
    \hline
    \textit{Treatment: Feedback} & & & & & & & & &\\
    null model(random intercept) & 27 & 48097 & 48303 & -24021 & 48043 &  &  &  & \\
    + random slopes & 29 & 48100 & 48321 & -24021 & 48042 & 1 & 2 & 0.67 & 1.00\\
    \hline
    \multicolumn{8}{l}{The null model is model \ref{si-spec:re} with $b_q$ and without $\gamma_q$;}\\
    \multicolumn{8}{l}{the alternative model is model \ref{si-spec:re}.}\\
    \end{tabular}%
    }
    \label{si-tab:trust_slope}
\end{table}

\begin{table}[htbp!]
\caption{Testing Treatment Effects Heterogeneity on Willingness to Share}
\centering
\makebox[\linewidth][c]{%
\begin{tabular}[t]{lrrrrrrrrr}
    \hline
     & npar & AIC & BIC & logLik & deviance & $\chi^2$ & Df & $P(>\chi^2)$ & adj.p\\
     \hline
    \textit{All treatments included} & & & & & & & & &\\
    null model(random intercept)& 31 & 173907 & 174177 & -86923 & 173845&  &  &  & \\
    + random slopes & 51 & 173885 &174329 &-86891 & 173783 & 63 & 20 & 0.00 & 0.00\\
    \hline
    \textit{Treatment: Traditional search} & & & & & & & & &\\
    null model(random intercept)& 27 & 58491 & 58697 & -29219 & 58437&  &  &  & \\
    + random slopes & 29 & 58492 & 58713 & -29217 & 58434 & 3 & 2 & 0.22 & 0.65\\
    \hline
    \textit{Treatment: Explanation} & & & & & & & & &\\
    null model(random intercept) & 27 & 57822 & 58027 & -28884 & 57768  &  &  &  & \\
    + random slopes & 29 & 57826 & 58046 & -28884 & 57768 & 0 & 2 & 1.00 & 1.00\\
    \hline
    \textit{Treatment: Reference} & & & & & & & & &\\
    null model(random intercept) & 27 & 56501 & 56706 & -28224 & 56447  &  &  &  & \\
    + random slopes& 29 & 56500 & 56720 & -28221 & 56442 & 5 & 2 & 0.09 & 0.38\\
    \hline
    \textit{Treatment: Highlighting} & & & & & & & & &\\
    null model(random intercept) & 27 & 57689 & 57894 & -28818 & 57635 &  &  &  & \\
    + random slopes & 29 & 57676 & 57896 & -28809 & 57618 & 17 & 2 & 0.00 & 0.00\\
    \hline
    \textit{Treatment: Feedback} & & & & & & & & &\\
    null model(random intercept) & 27 & 59011 & 59217 & -29478 & 58957  &  &  &  & \\
    + random slopes & 29 & 59014 & 59235 & -29478 & 58956 & 1 & 2 & 0.78 & 1.00\\
    \hline
    \multicolumn{8}{l}{The null model is model \ref{si-spec:re} with $b_q$ and without $\gamma_q$;}\\
    \multicolumn{8}{l}{the alternative model is model \ref{si-spec:re}.}\\
    \end{tabular}%
    }
    \label{si-tab:share_slope}
\end{table}

\newpage

\section{Randomization Inference}
We conducted randomization inference for between-subjects and within-subjects effects. We tested the sharp null of no individual-level treatment effects ($H_0 : \beta_{ti}=0$ for any individual i, between-subjects or within-subjects treatment t) and implemented randomization inference by randomly redistributing the experiment group assignment. The results are reported in Table \ref{si-tab:ri-btw} and \ref{si-tab:ri-wi}. Except for the traditional search effect on 5-item trust, all the other significance levels of the primary analysis results remained unchanged with randomization inference. 
\begin{table}[htbp!]
   \caption{Randomization Inference of Between-subjects Treatment Effects}
   \begin{minipage}{\textwidth}
      \centering
   \begin{tabular}{lcc} 
   \midrule
                  & 5-item Trust  & Willingness to Share\\ 
      \\
      \emph{Treatment}\\
      Traditional search                   & 0.043 (0.272)         & 0.141 (0.024)\\      
      Explanation              & -0.011 (0.766)        & 0.011 (0.852)\\    
      Reference                 & 0.091 (0.030)  & 0.121(0.044)\\
      Uncertainty Highlighting & -0.157 (0.00) & -0.179 (0.002)\\  
      Social Feedback          & 0.018 (0.640)       & 0.022 (0.676)\\    
      \\
      Covariate Controls       & Yes            & Yes\\  
      \midrule
      \emph{Fixed Effects}\\
      search task                    & Yes            & Yes\\  
      \midrule
      \multicolumn{3}{l}{Results from 500 simulations;}\\
      \multicolumn{3}{l}{Randomization inference p-values in parentheses;}\\
   \end{tabular}
   \label{si-tab:ri-btw}
   \end{minipage}
\end{table}

\begin{table}[htbp!]
    \caption{Randomization Inference of Within-subjects Effects}
    \begin{minipage}{\textwidth}
        \centering
        \begin{tabular}{lccc}
        \midrule
              & Reference & Highlighting & Feedback\\
            \midrule
            \addlinespace[0.5em]
            \multicolumn{4}{l}{\textit{Outcome: 5-item Trust}}\\
            \midrule \hspace{1em}variation & -0.021(0.408) & -0.338(0.00) & -0.213(0.00)\\
            \hspace{1em}FE: Individual & Yes & Yes & Yes\\
            \hspace{1em}FE: task & Yes & Yes & Yes\\
            \addlinespace[0.5em]
            \multicolumn{4}{l}{\textit{Outcome: Willingness to Share}}\\
            \midrule \hspace{1em}variation& -0.008 (0.820) & -0.490(0.00) & -0.286(0.00)\\
            \hspace{1em}FE: Individual & Yes & Yes & Yes\\
            \hspace{1em}FE: task & Yes & Yes & Yes\\
            \midrule\midrule
            \multicolumn{3}{l}{Results from 500 simulations;}\\
            \multicolumn{3}{l}{Randomization inference p-values in parentheses;}\\
        \end{tabular}
        \label{si-tab:ri-wi}
    \end{minipage}
\end{table}

\newpage

\section{Placebo Tests}
Although the GenAI, traditional search, and explanation groups did not have any variations, the treatment variation assignment algorithm was still implemented in these groups. As participants were randomly assigned to identical baseline and variation treatments in these groups, we should observe no differences in their effects. As expected, none of the three groups showed significant differences, lending credibility to the randomization and to the treatment effects we observed in the other groups (see Table \ref{si-tab:placebo}). 

\begin{table}[htbp!]
    \caption{Placebo Test of Within-subjects Effects in GenAI, Traditional Search, and Explanation Groups}
    \begin{minipage}{\textwidth}
        \centering
        \begin{tabular}{lccc}
        \midrule
              & GenAI & Traditional search & Explanation\\
            \midrule
            \addlinespace[0.5em]
            \multicolumn{4}{l}{\textit{Outcome: 5-item Trust}}\\
            \midrule 
            \hspace{1em}variation & -0.012 & -0.012 & -0.028\\
            & (0.014) & (0.022) & (0.024)\\
            \hspace{1em}FE: Individual & Yes & Yes & Yes\\
            \hspace{1em}FE: task & Yes & Yes & Yes\\
            \hspace{1em}Observations & 7326 & 7578 & 7308\\
            \addlinespace[0.5em]
            \multicolumn{4}{l}{\textit{Outcome: Willingness to Share}}\\
            \midrule \hspace{1em}variation& -0.020 & -0.009 & -0.017\\
             & (0.019) & (0.033) & (0.022)\\
            \hspace{1em}FE: Individual & Yes & Yes & Yes\\
            \hspace{1em}FE: task & Yes & Yes & Yes\\
            \hspace{1em}Observations & 7326 & 7578 & 7308\\
            \midrule\midrule
            \multicolumn{3}{l}{\emph{Two-way clustered standard-errors in parentheses}}\\
            \multicolumn{3}{l}{\rule{0pt}{1em}* p $<$ 0.05, ** p $<$ 0.01, *** p $<$ 0.001}\\
        \end{tabular}
        \label{si-tab:placebo}
    \end{minipage}
\end{table}

\clearpage
\bibliographystyle{unsrt}
\bibliography{sombib}

\end{document}